\documentclass[useAMS,usenatbib]{mn2e}

\title[Radio structures of the nuclei of nearby Seyfert galaxies]{Radio
  structures of the nuclei of nearby Seyfert galaxies and the nature
  of the missing diffuse emission}
\author[M. Orienti \& M.A. Prieto]
  {M. Orienti$^{1,2}$\thanks{E-mail: orienti@ira.inaf.it},
M.A. Prieto$^1$\\
$^1$Instituto de Astrof\'{i}sica de Canarias, c/ V\'{i}a L\'actea s/n,
E-38205  La Laguna (Tenerife), Spain\\
$^{2}$Istituto di Radioastronomia - INAF, Via P. Gobetti 101, I-40129 Bologna, Italy}

\date{Received \today; accepted ?}

\pagerange{\pageref{firstpage}--\pageref{lastpage}} \pubyear{2002}

\def\LaTeX{L\kern-.36em\raise.3ex\hbox{a}\kern-.15em
    T\kern-.1667em\lower.7ex\hbox{E}\kern-.125emX}

\begin{document}

\label{firstpage}

\maketitle

\begin{abstract}
We present archival high spatial resolution VLA and VLBA data 
of the nuclei of seven of
the nearest and brightest Seyfert galaxies in the Southern Hemisphere. 
At VLA resolution ($\sim 0.1$ arcsec), the nucleus of
the Seyfert galaxies is unresolved, with the exception of
MCG-5-23-16 and NGC\,7469 showing a core-jet structure. 
Three Seyfert nuclei are surrounded by diffuse radio emission related
to star-forming regions.
VLBA observations with parsec-scale resolution 
pointed out that 
in MRK\,1239 the nucleus is clearly resolved in
two components separated by $\sim$ 30 pc, while the nucleus of NGC\,3783 
is unresolved.
Further comparison between VLA and VLBA data of 
these two sources
shows that the flux density
at parsec scales is only 20\% of that measured by the VLA. This suggests
that the radio emission is not concentrated in a
single central component, as in elliptical radio galaxies, and an 
additional low-surface brightness
component must be present. 
A comparison between Seyfert nuclei with different
radio spectra points out that the ``presence'' of undetected flux on
milli-arcsecond scale is common in steep-spectrum objects, while in
flat-spectrum objects essentially all the radio emission is
recovered. In the steep-spectrum objects, 
the nature of this ``missing'' flux is likely due
to non-thermal AGN-related radiation, perhaps from a jet that gets
disrupted in Seyfert galaxies because of the denser environment of
their spiral hosts. 

\end{abstract}


\section{Introduction}

Only a small fraction
($\sim$ 10\%) of the population of 
active galactic nuclei (AGN) possess a powerful
radio emission ($L_{\rm 1.4\,GHz} > 10^{23}$ W/Hz), 
as found in radio galaxies/quasars and blazars.
Seyfert galaxies are part of the ``radio-quiet''
AGN population, with radio luminosity $L_{\rm 1.4\,GHz} \leq 10^{20 -
  23}$ W/Hz.  
Despite their weak radio emission, Seyfert nuclei are
very nearby, allowing the study in detail of the
radio properties of their central engine.\\  
Radio observations with arcsecond resolution of several Seyfert
samples \citep[see e.g.][]{ulvestad84, morganti99, thean00} showed that
a large fraction of Seyferts have resolved structures, with hint
of jets and/or extended emission, the latter usually related to 
star-forming regions.
Several objects, such as NGC\,1052
\citep{wrobler84}, NGC\,1068 \citep{ulvestad87}, 
NGC\,7674 \citep{momjian03}, and MRK\,3 \citep{kukula99}
have been found to display a radio morphology with
core, collimated jets and hot spots, similar
to those found in radio-loud galaxies. 
However, powerful radio sources have
linear structures reaching hundreds of kpc or even Mpc scales, while
in Seyferts the radio emission is confined to a few kpc or 
even sub-kpc scales.\\
When observed with milli-arcsecond resolution, the pc-scale structure
of Seyfert nuclei is usually resolved in several components
\citep[e.g. NGC\,3079,][]{trotter98}, resembling a jet
structure \citep[e.g. NGC\,4151,][]{ulvestad98, nagar01}
and sometimes with the presence of extended emission
\citep[e.g. NGC\,5793,][]{haghi00}. 
The comparison between arcsec and
milli-arcsec radio properties pinpointed a frequent misalignment
between pc and kpc-scale jets, suggesting either a change in jet
ejection axis, or a bending due to pressure gradients in the
ambient medium \citep{middelberg04}. \\
An intriguing characteristic shown by a large number of Seyfert
nuclei 
is that the radio emission 
arising from their pc-scale structure is often much fainter than
that derived from observations with lower resolution, even in the case the
nucleus is unresolved. This result suggests that in Seyfert nuclei
the radio emission is not concentrated in the central region, as
found in powerful radio galaxies, but it
extends on scales of tens or hundreds pc \citep[see
e.g.][]{sadler95}. 
However, not all the Seyfert nuclei have missing flux on parsec
scales, as in the case of MRK\,530 \citep{lal04}, indicating that the
radio emission mainly arises from the central compact component,
without evidence of extended, low-surface brightness features.\\
In this 
paper, we present the results of multi-frequency archival VLA and/or VLBA 
data of a sample of some of the nearest and brightest Seyfert
galaxies, taken from the infrared high spatial resolution studies
conducted by \citet{prieto09} and \citet{reunanen09}.
For these sources, no
complete and/or unambiguous 
information on their radio properties could be found in the
literature. 
The comparison between
these data with those at pc-scale resolution 
available from the literature allows a better
determination of the physical conditions of the radio emission at kpc
and pc scales. 

\begin{figure*}
\begin{center}
\includegraphics{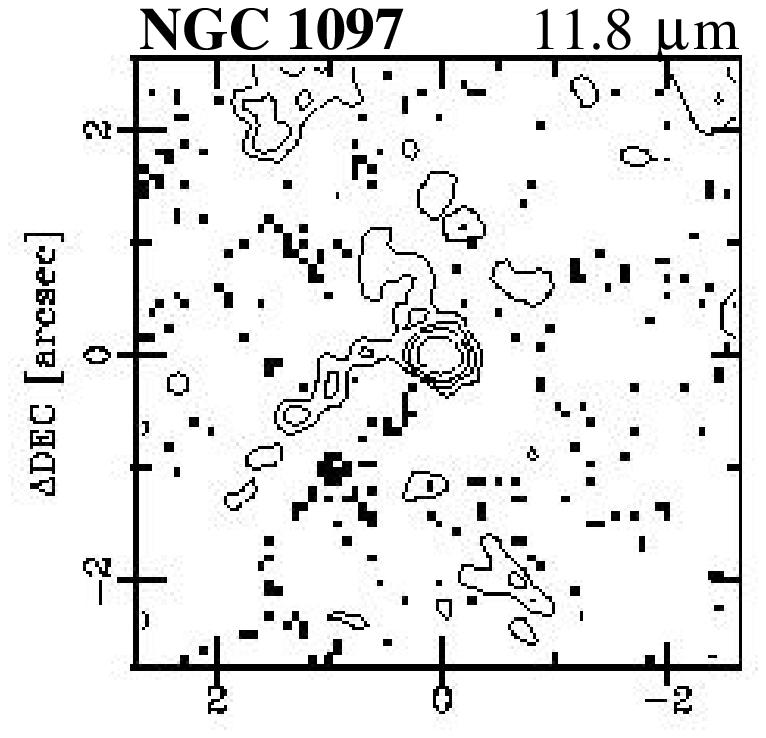}
\includegraphics{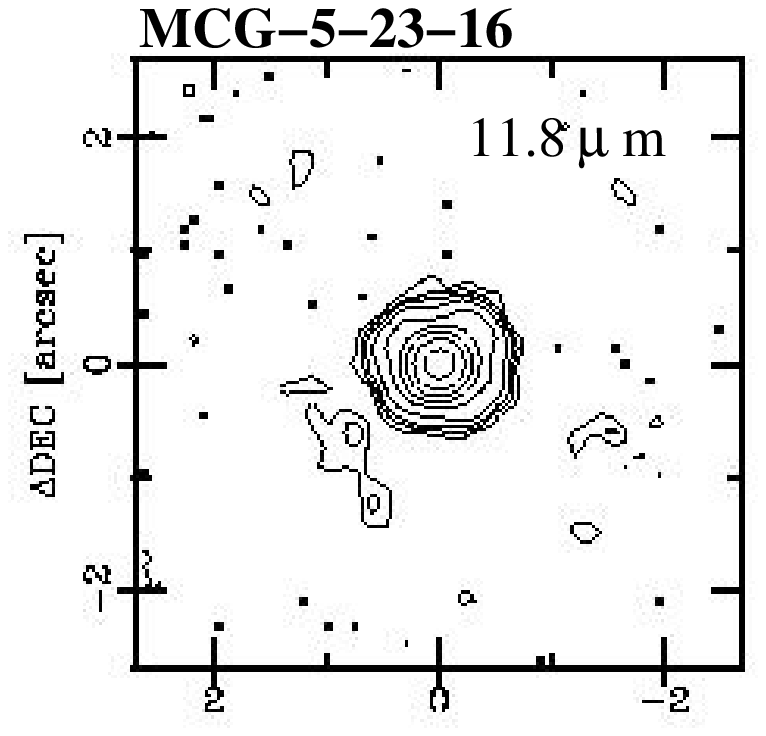}
\includegraphics{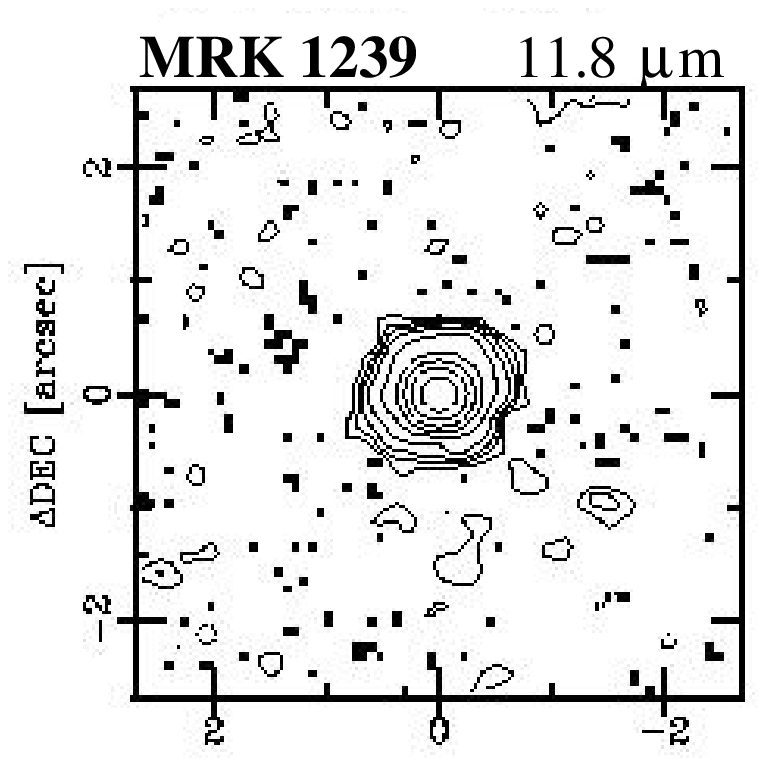}
\includegraphics{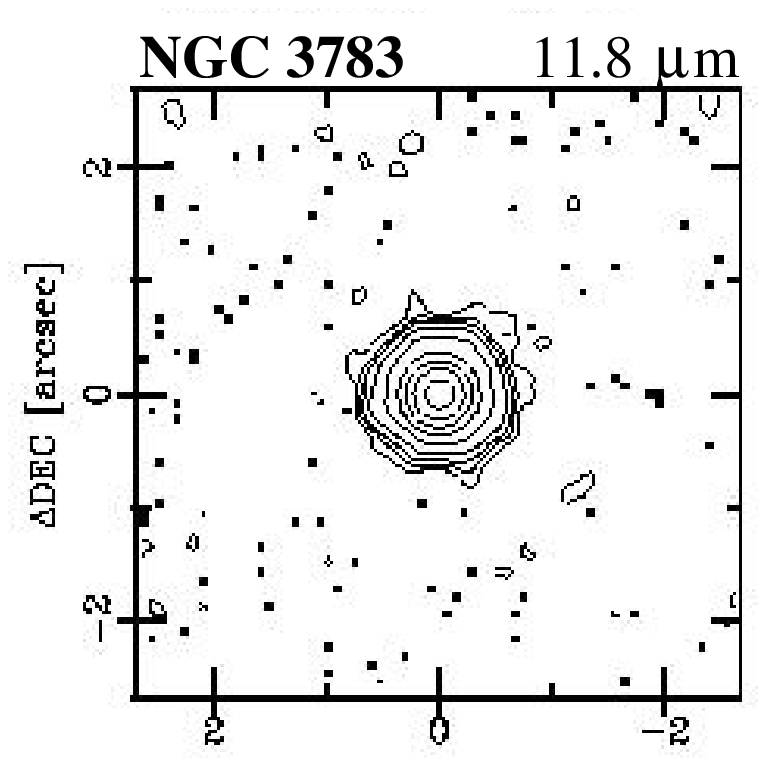}
\includegraphics{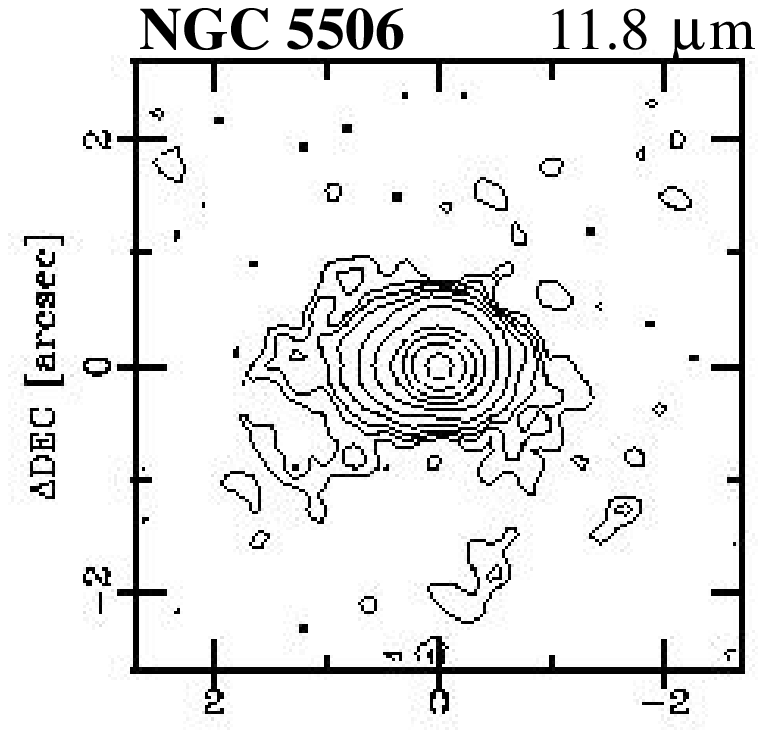}
\includegraphics{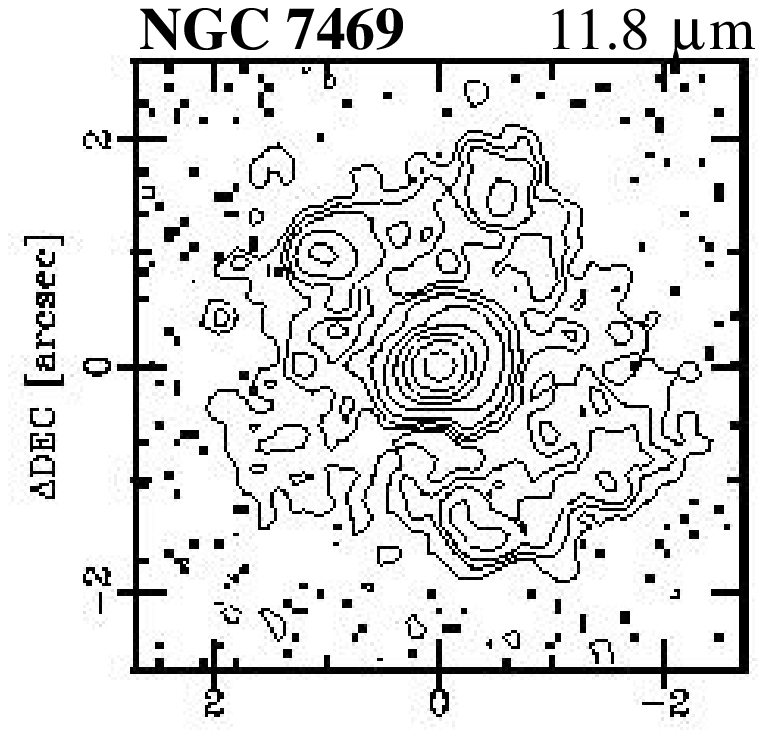}
\includegraphics{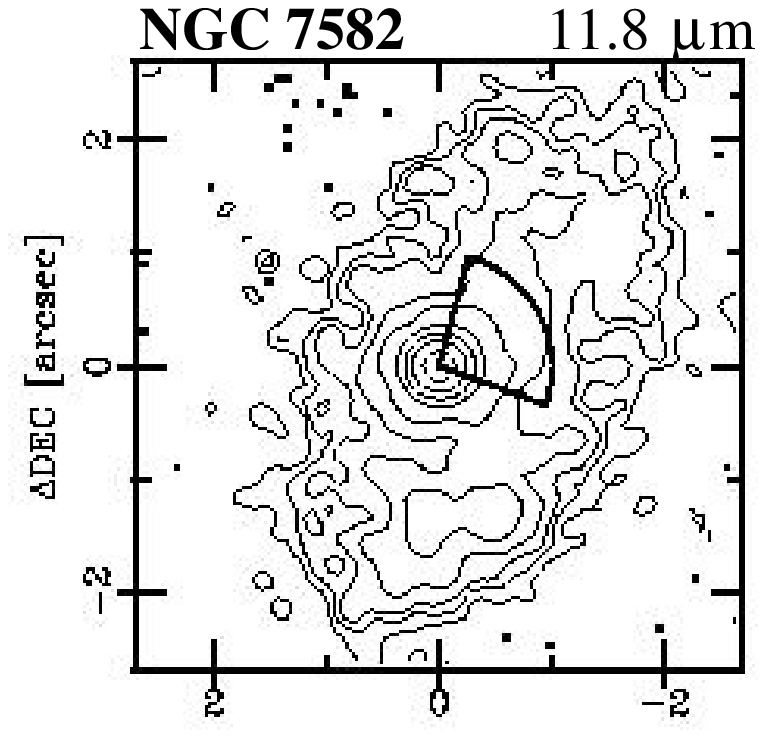}
\vspace{18cm}
\caption{VLT-VISIR images at 11.8 $\mu$m of the Seyfert nuclei studied
in this paper. Images adapted from \citet{reunanen09}. Contours
are at 3$\sigma$, 5, 7, 11 and 19 levels. The triangle in NGC\,7582
represents the ionization cone.}
\label{ir}
\end{center}
\end{figure*}

Throughout this paper, we assume the following cosmology: $H_{0} =
71\, {\rm km\, s^{-1}\, Mpc^{-1}}$, $\Omega_{\rm M} = 0.27$, and
$\Omega_{\Lambda} = 0.73$, in a flat Universe. The spectral index
is defined as 
$S {\rm (\nu)} \propto \nu^{- \alpha}$.

\section{The sample}

The objects studied in this paper are part of the sample of
Seyfert nuclei from \citet{prieto09} and \citet{reunanen09}. They
consist in the nearest
and brightest Seyfert nuclei accessible from the Southern
Hemisphere.
The proximity of these objects allows us to study them at scales of a
few tens of parsec in the IR with 
VLT - adaptive optics and
interferometry which deliver spatial resolutions comparable to that
achieved in the optical with HST, and in radio with VLA and VLBI. 
In
Fig. \ref{ir} we present VLT-VISIR images at 11.8 $\mu$m of the
Seyfert studied in this paper. Details on VISIR observations and data
reduction can be found in \citet{reunanen09}.
The dominant IR emission comes from the
central nucleus that is unresolved with FWHM $<$ 0.35 arcsec, which
corresponds to linear sizes between 30 and 190 pc, depending on the
redshift. In NGC\,7469 and NGC\,7582, an additional emission from the star
forming regions is clearly detected.\\
High spatial resolution spectral energy distribution (SED) have been
produced for these objects.
For the majority of the Seyfert considered,
multi-frequency, high-resolution information in the radio band has not
been exploited in full. For
this reason, we
collected and analysed archival radio data with (sub-)arcsecond resolution 
in order to implement the information already available
and to constrain
the radio emission from these nuclei. \\
The list of the Seyfert nuclei
studied in this paper is in Table \ref{sample}. We considered only
datasets at frequencies and resolution not available in the
literature, or for which no unambiguous information could be found.\\

\subsection{Radio properties of the selected sources}

\subsubsection{NGC\,1097}

NGC\,1097 is a barred spiral
galaxy hosting a Seyfert 1 nucleus \citep{storchi93}.
The radio emission 
is dominated by a ring-like structure related to star-forming regions
also detected in optical.
At the centre of the ring there is an unresolved (beam $\sim$ 0.3
arcsec at 8.4 GHz)
component \citep{thean00}
with an inverted non-thermal spectrum \citep{morganti99}.    
The radio core was observed by \citet{sadler95}
with the Parkes-Tidbinbilla interferometer
(PTI) that provides a resolution $<$0.1 arcsec.
No radio emission was detected, implying that 
the core flux density is $<$5 mJy at
2.3 GHz.\\

\subsubsection{MCG-5-23-16}

MCG-5-23-16 is a S0 galaxy hosting a Seyfert 2 nucleus. 
Its radio properties were
studied by \citet{ulvestad84} with VLA observations at 1.4 and
4.8 GHz. The radio structure of the nucleus is found slightly resolved
(beam $\sim$ 0.4 arcsec at 4.8 GHz)
with a spectral index  $\alpha \sim 0.5$. 
No evidence of extra-nuclear emission 
was detected. \\
High spatial resolution (beam $<$ 0.1 arcsec) observations with PTI
\citep{sadler95} could set only an upper limit to the flux density of
the core $S_{\rm 8.4 GHz} <$7 mJy.\\

\subsubsection{MRK\,1239}

MRK\,1239 is an early-type E/S0 galaxy hosting a Seyfert 1.5 nucleus.
Its central radio structure was
studied by \citet{ulvestad95} and \citet{rush96} with the VLA. 
The source appears unresolved at all frequencies (beam $\sim$ 0.2
arcsec at 8.4 GHz) 
without evidence
of extended emission surrounding the nucleus. \citet{ulvestad95} noted
a strong steepness in the radio spectrum between 4.8 and 8.4 GHz ($\alpha
\sim 1.6$), that they interpreted as due to 
flux variability
(observations were not carried out simultaneously). \\

\subsubsection{NGC\,3783}

NGC\,3783 is a barred spiral galaxy with a
highly variable Seyfert 1 nucleus. Multi-frequency radio VLA observations 
by \citet{ulvestad84}, \citet{unger86} and \citet{schmitt01} pointed out the
presence of an unresolved component (beam $\sim$0.25 arcsec at 8.4
GHz) with a spectral index $\alpha \sim
0.5$. No extended extra-nuclear emission was found. 
High spatial resolution (beam $<$ 0.1 arcsec) observations with PTI 
\citep{sadler95} could set only an upper limit to the flux density of
the core $S_{\rm 1.6 GHz} <$6 mJy.\\

\begin{figure}
\begin{center}
\includegraphics{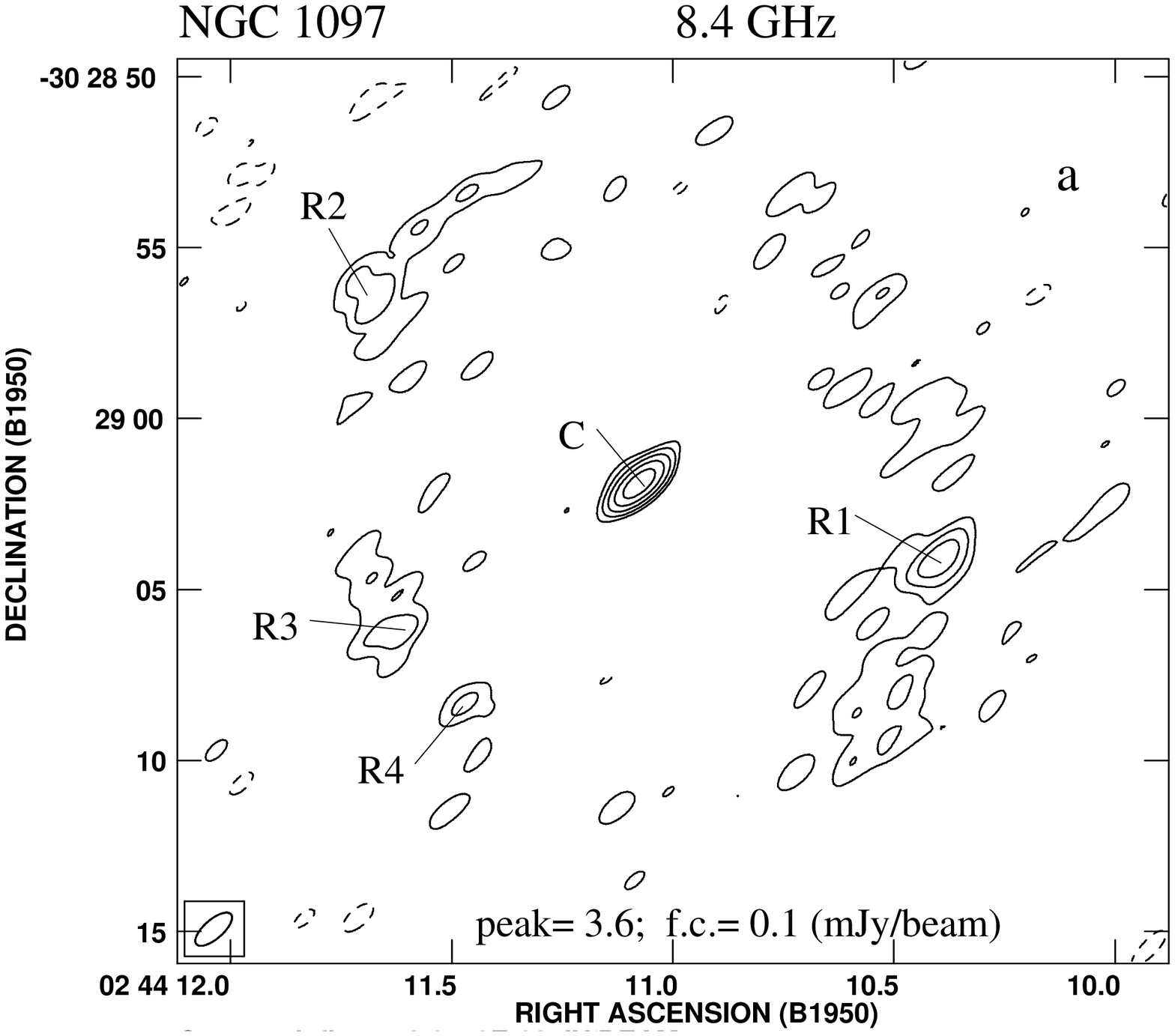}
\includegraphics{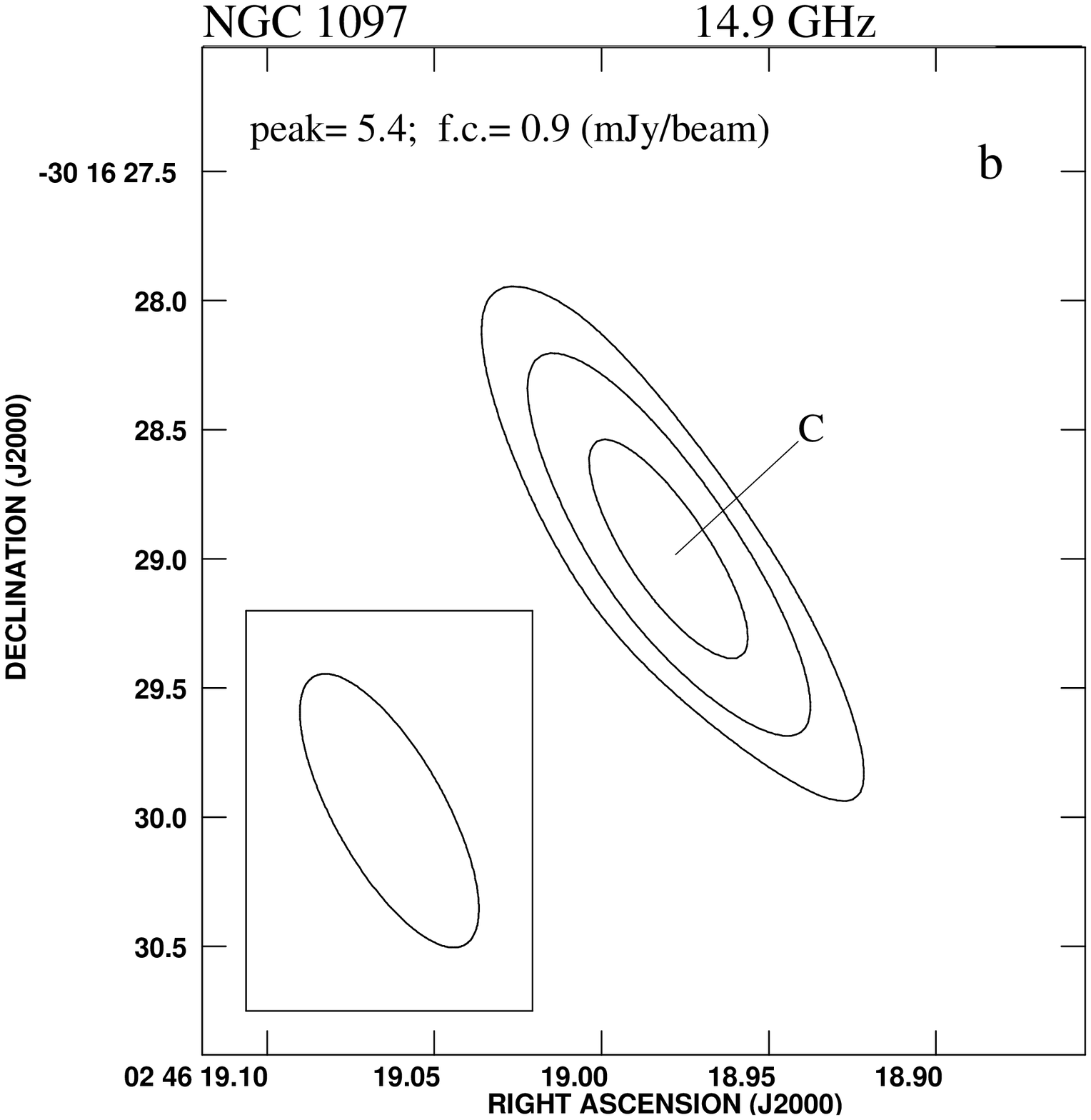}
\vspace{14.5cm}
\caption{VLA images at 8.4 GHz ({\it top}) and at 14.9 GHz ({\it
    bottom}) of the central region of NGC\,1097. On each image we
    provide the observing frequency; the restoring beam, plotted on
    the bottom left corner; the peak flux density in
    mJy/beam; the first contour intensity ({\it f.c.} in mJy/beam),
    that is 3 times the off-source noise level; contour levels
    increase of a factor 2.
    }
\label{n1097}
\end{center}
\end{figure}

\begin{figure}
\begin{center}
\includegraphics{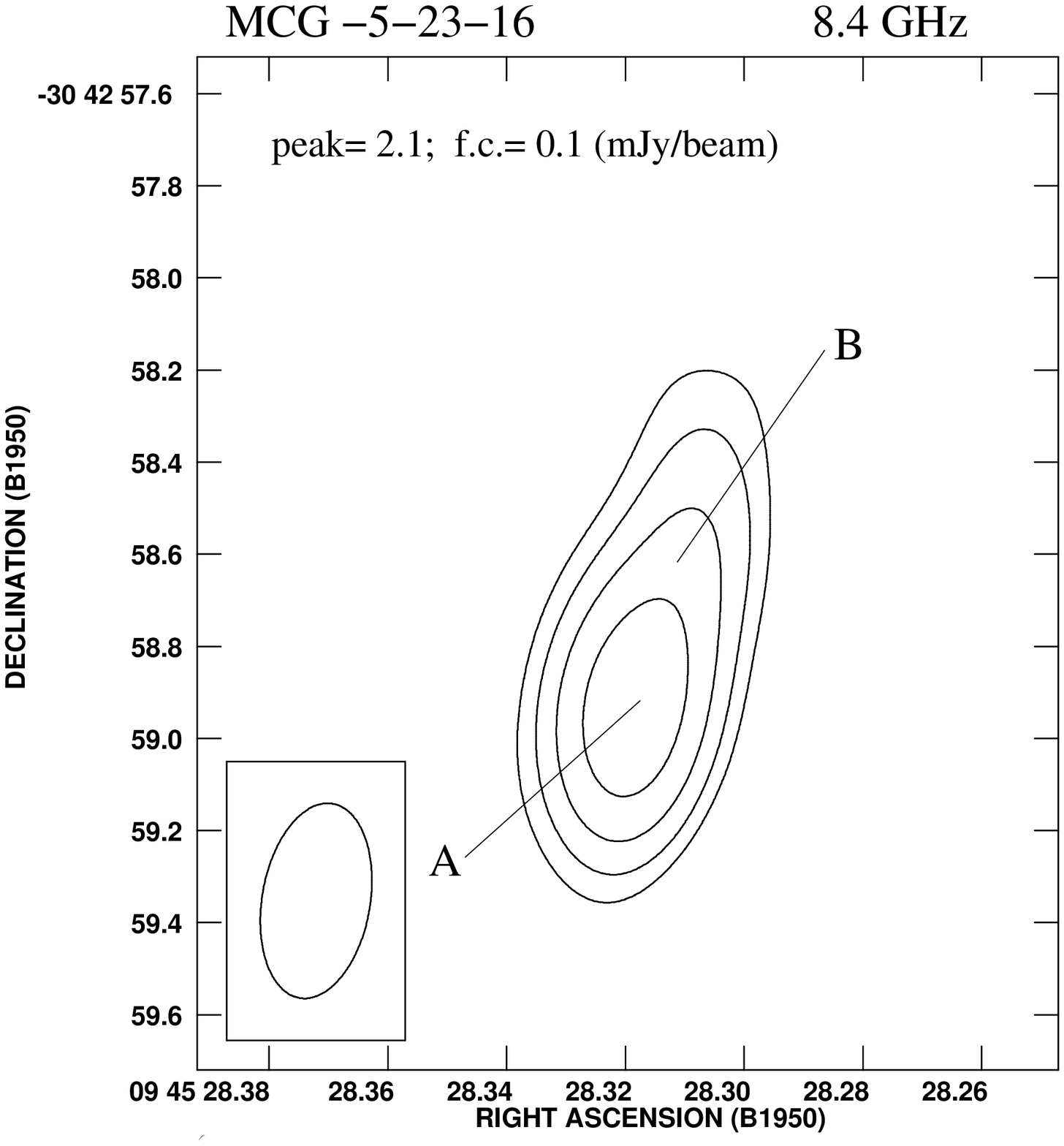}
\vspace{8.0cm}
\caption{VLA image at 8.4 GHz of the central region of MCG-5-23-16. 
On the image we
    provide the observing frequency; the restoring beam, plotted on
    the bottom left corner; the peak flux density in
    mJy/beam; the first contour intensity ({\it f.c.} in mJy/beam),
    that is 3 times the off-source noise level; contour levels
    increase of a factor 2.}
\label{m-5}
\end{center}
\end{figure}

\subsubsection{NGC\,5506}

NGC\,5506 is a spiral galaxy with a Seyfert 2 nucleus. 
The central radio structure was studied by
\citet{ulvestad81} and \citet{unger86} by VLA and MERLIN
observations. It shows 
a diffuse bubble-like feature extending to the north-west of the
unresolved nucleus. At 5 GHz \citet{wehrle87} found a
low-surface brightness halo enshrouding the central structures.\\
Multi-frequency observations of the nucleus with the PTI 
\citep{sadler95} showed that it
has a convex spectrum. 
Pc-scale VLBI observations at 18, 6 and 3.6 cm \citep{middelberg04}
could resolve the nucleus in three main components roughly aligned
in the east-west direction and extending within an area of about 50
mas ($\sim$ 6 pc). The flattish spectrum displayed by the brightest
component suggests that it is the true source core. \\

\subsubsection{NGC\,7469}

NGC\,7469 is a spiral galaxy with a Seyfert
1 nucleus embedded in a ring of starburst activity. 
At radio frequencies, its structure shows an unresolved central component
surrounded by a ring of star-forming regions
\citep{wilson91}. MERLIN observations with high spatial resolution
(beam $\sim$ 0.05 arcsec) 
showed that the nucleus is clearly resolved in two components
with a core-jet structure elongated in the east-west
direction. VLBI observations (beam $\sim$ 10 mas)
at 1.6 GHz \citep{lonsdale03} could resolve the
core-jet structure in five different components, lying in an
east-west line as found in the lower resolution MERLIN image, and
contained in an area
of about 168 mas ($\sim$ 55 pc). \\

\subsubsection{NGC\,7582}

NGC\,7582 is a barred spiral galaxy hosting a Seyfert 2
nucleus surrounded by a star forming ring.
Its radio structure shows an unresolved
central component embedded in a diffuse emission related to the
star-forming region also detected in the optical
\citep{ulvestad84}. The extended emission is elongated in the
NW-SE direction, with 
a spectral index $\alpha  \sim 0.7$ \citep{morganti99}. 
High spatial resolution (beam $<$ 0.1 arcsec) observations with PTI
\citep{sadler95} could set only an upper limit to the flux density of
the core $S_{\rm 8.4 GHz} < 5$ mJy.\\

\begin{figure}
\begin{center}
\includegraphics{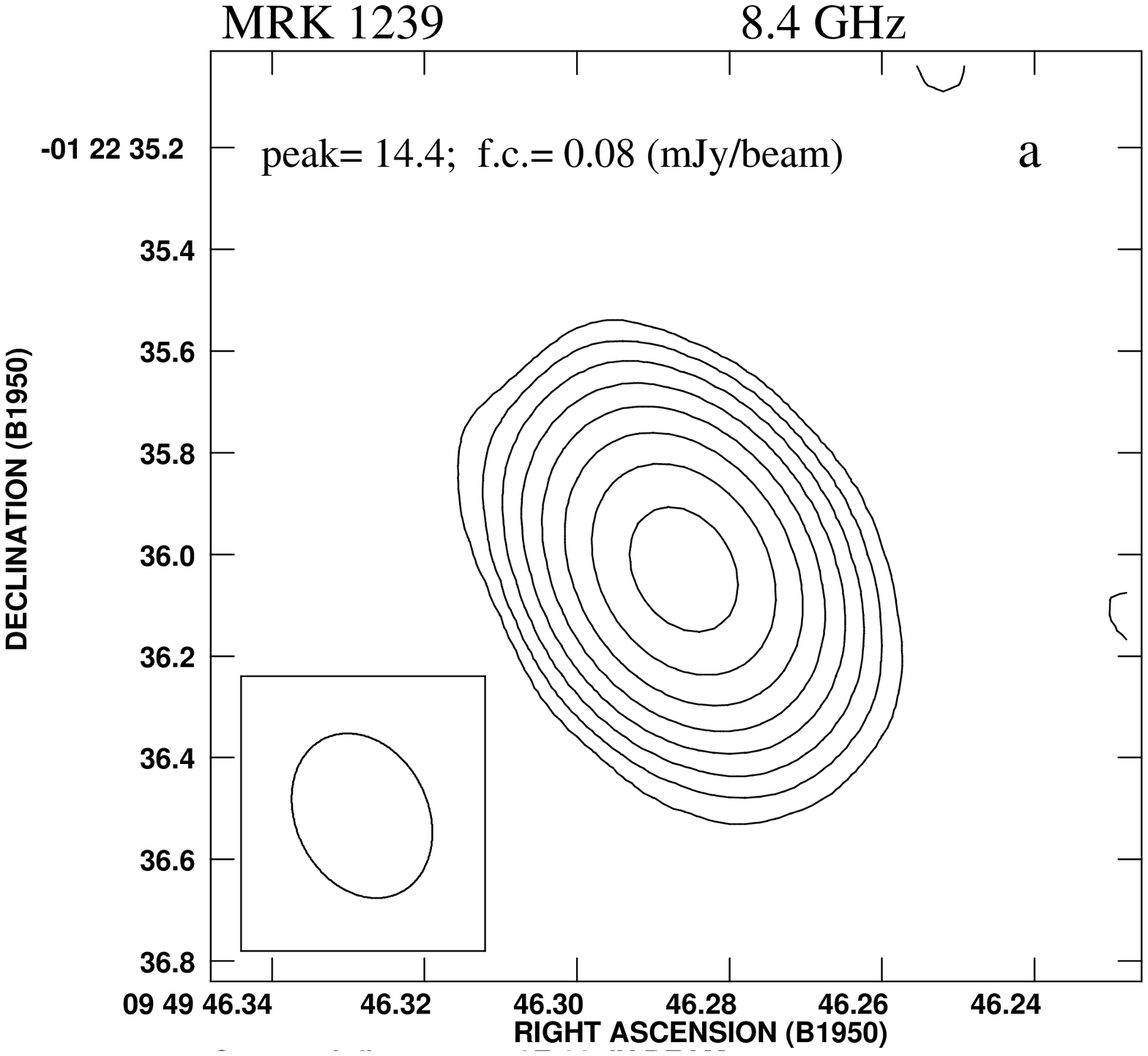}
\includegraphics{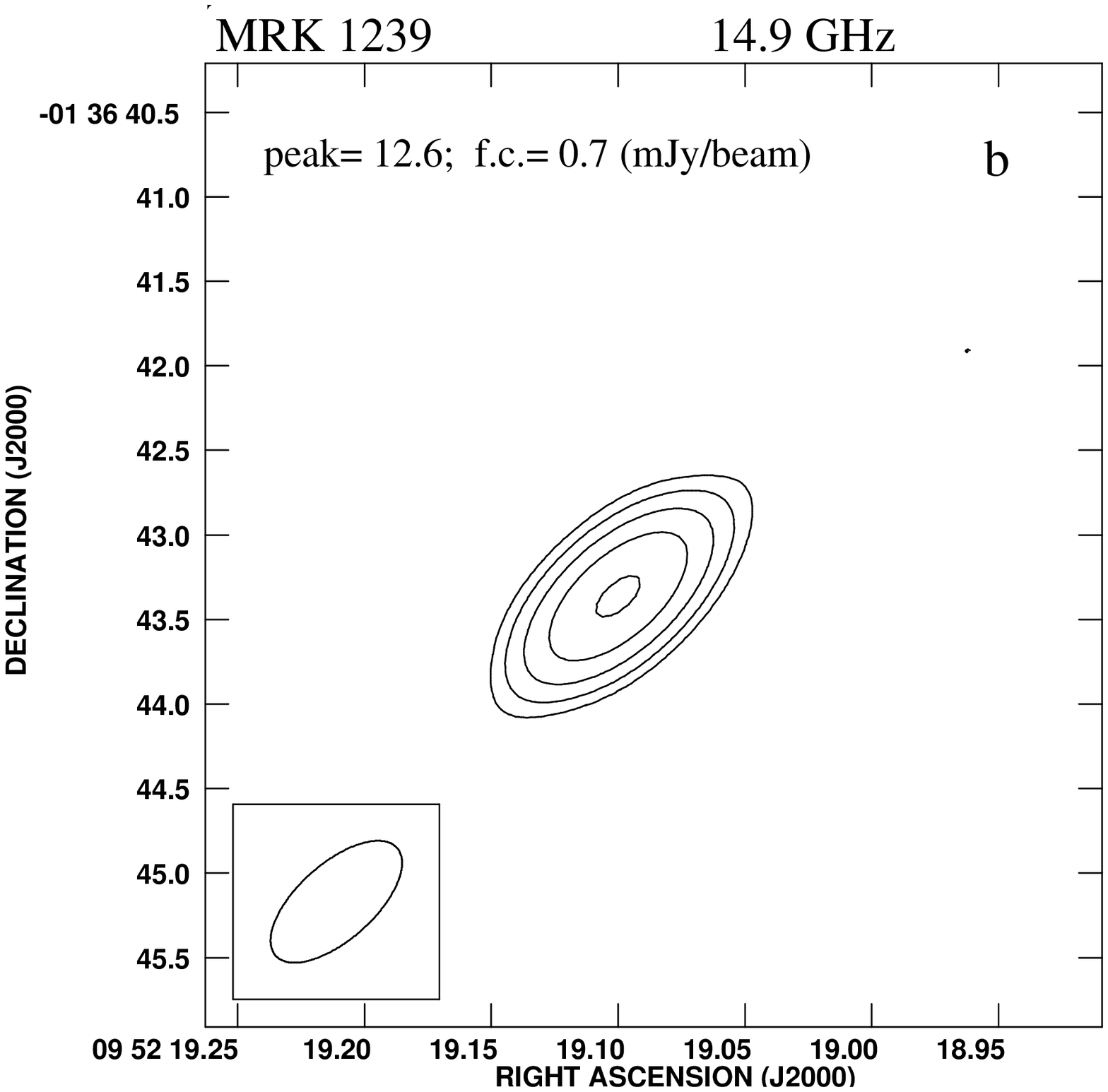}
\includegraphics{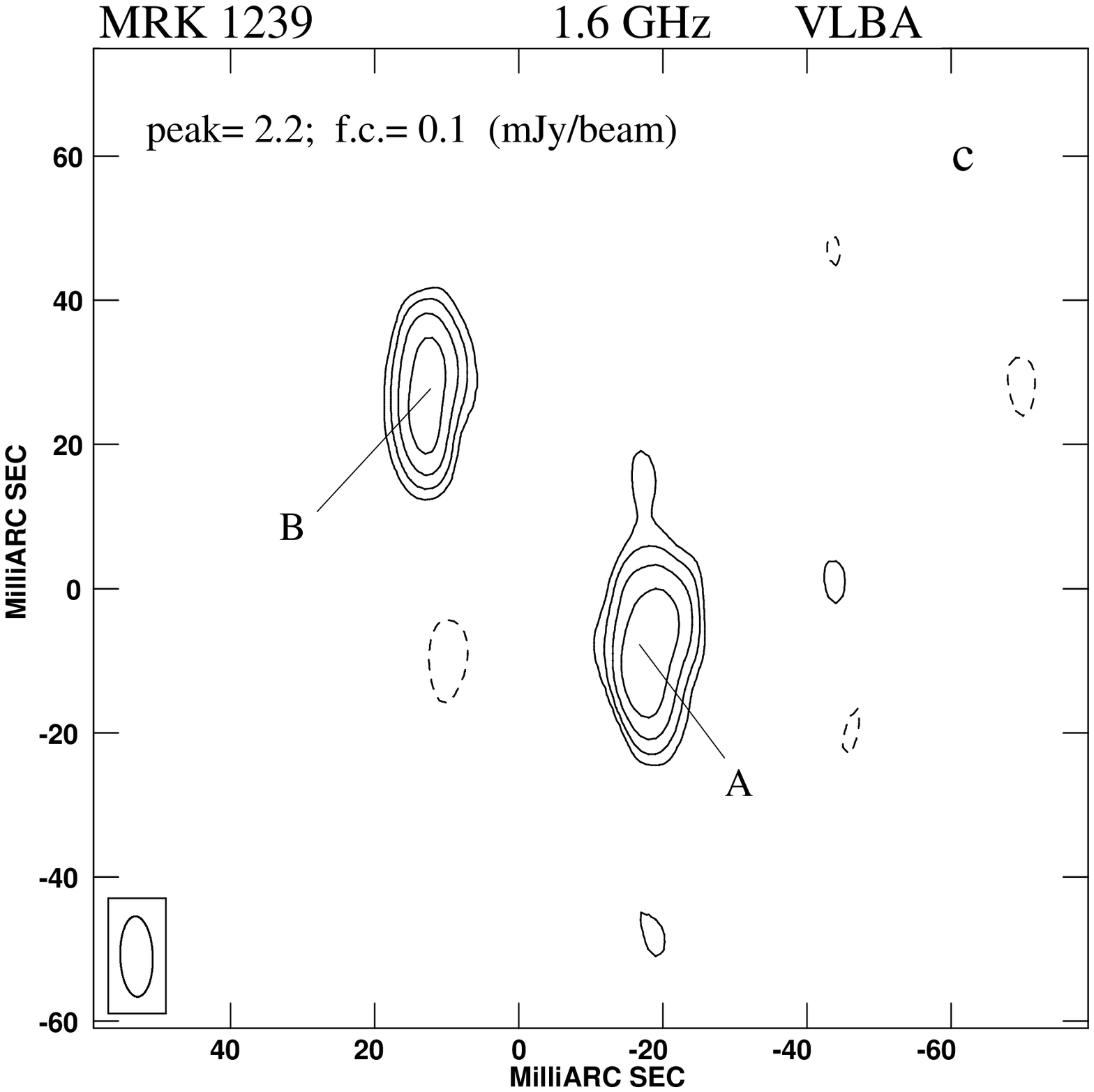}
\vspace{21.5cm}
\caption{VLA images at 8.4 GHz ({\it top}), and 14.9 GHz ({\it
    centre}), and VLBA image at 1.6 GHz ({\it bottom}) of 
the central region of MRK\,1239. On each image we
    provide the observing frequency; the restoring beam, plotted on
    the bottom left corner; the peak flux density in
    mJy/beam; the first contour intensity ({\it f.c.} in mJy/beam),
    that is 3 times the off-source noise level; contour levels
    increase of a factor 2.}
\label{m1239}
\end{center}
\end{figure}

\begin{figure}
\begin{center}
\includegraphics{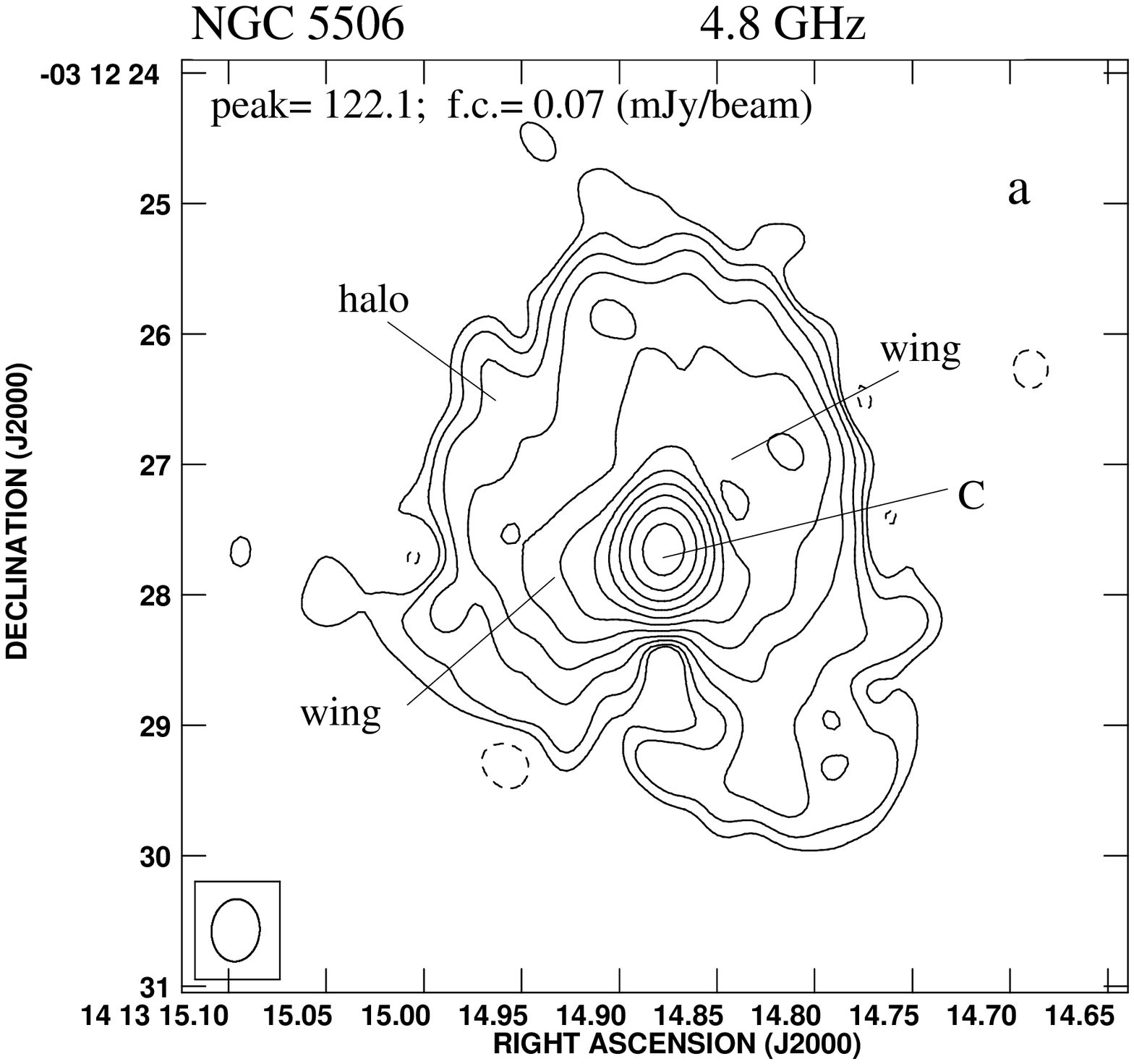}
\includegraphics{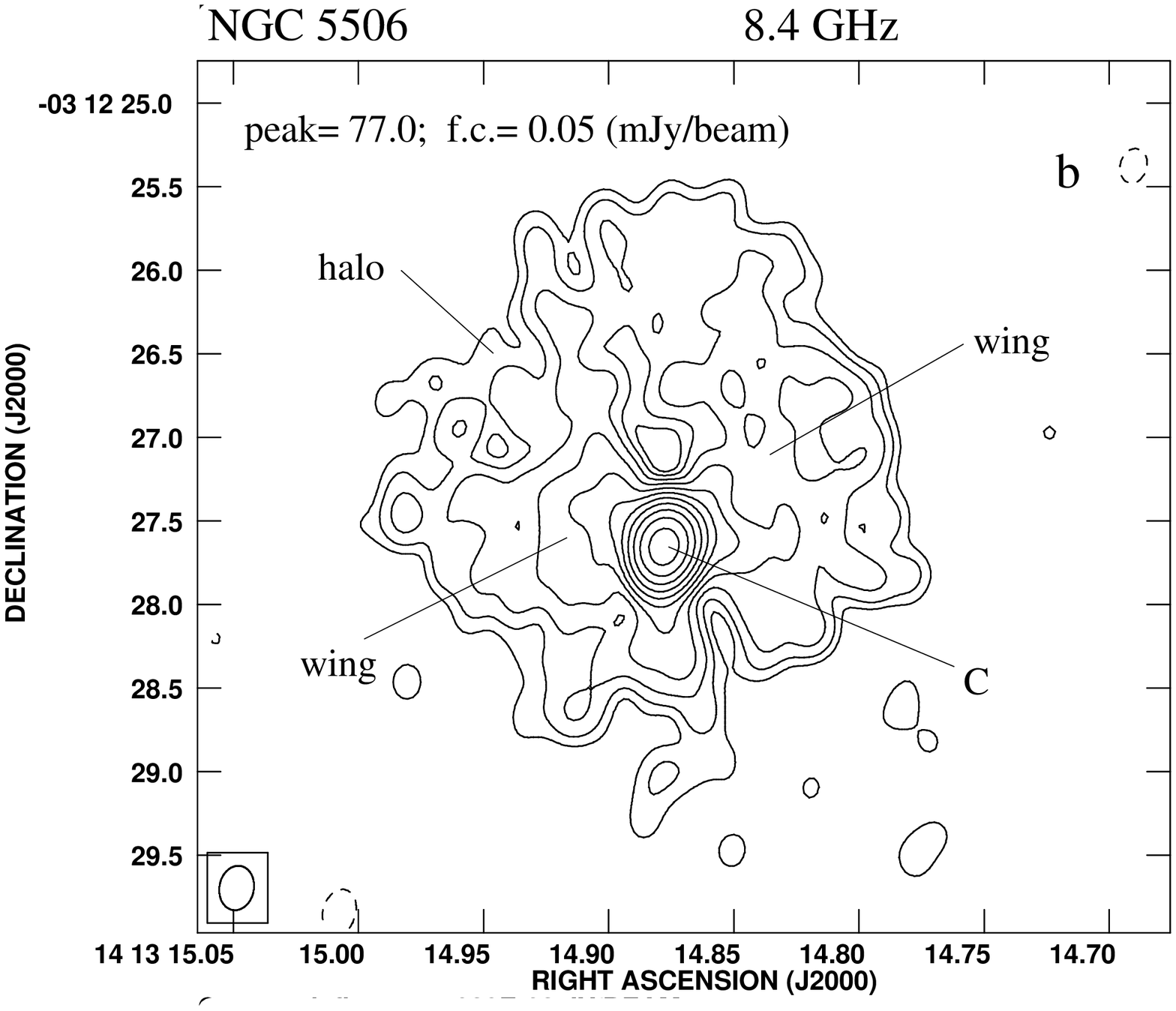}
\includegraphics{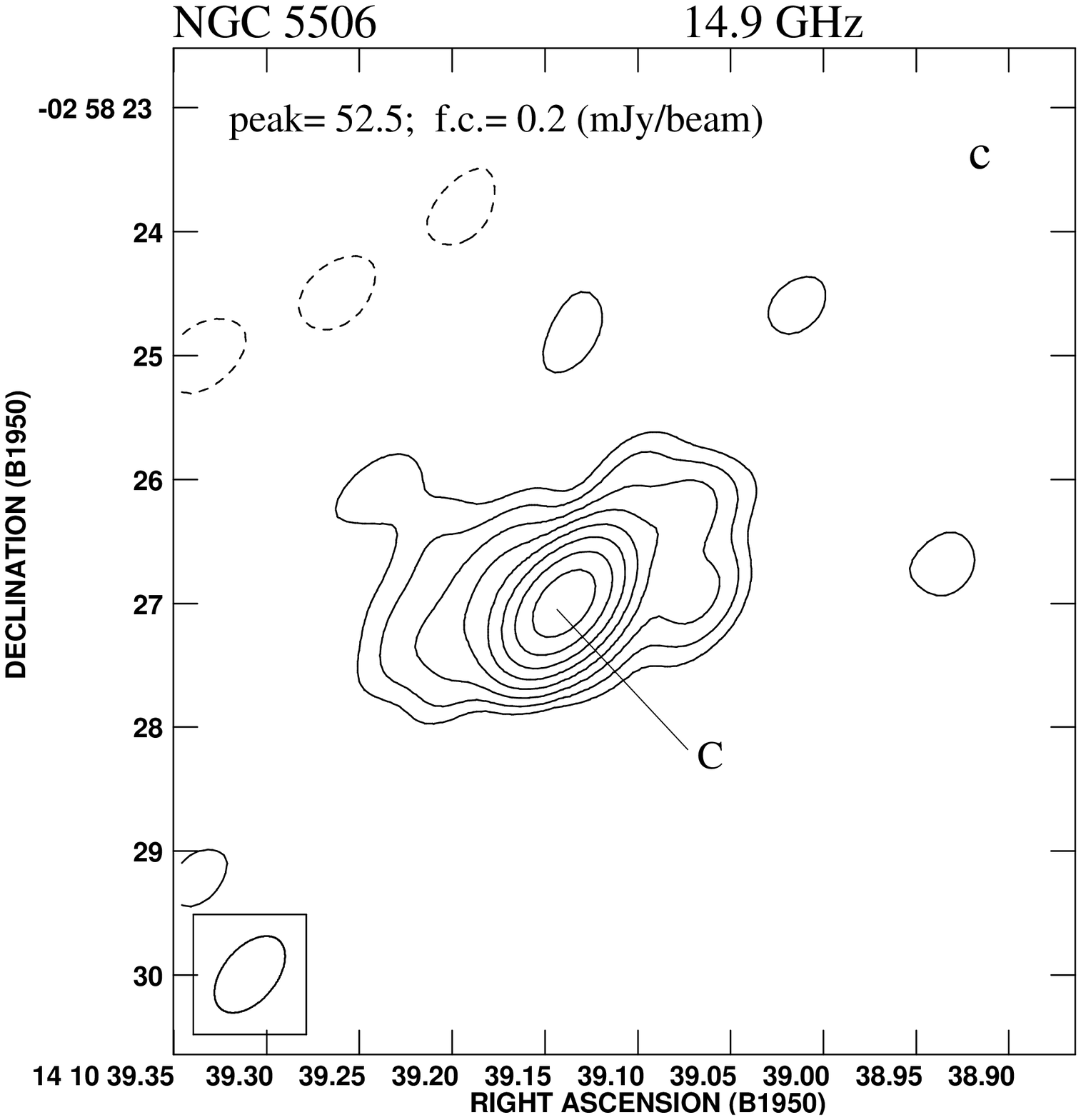}
\vspace{21.5cm}
\caption{VLA images at 4.8 GHz ({\it top}), 8.4 GHz ({\it
    centre}), and 14.9 GHz ({\it bottom}) of 
the central region of NGC\,5506. On each image we
    provide the observing frequency; the restoring beam, plotted on
    the bottom left corner; the peak flux density in
    mJy/beam; the first contour intensity ({\it f.c.} in mJy/beam),
    that is 3 times the off-source noise level; contour levels
    increase of a factor 2.}
\label{n5506}
\end{center}
\end{figure}

\begin{table}
\caption{The Seyfert sample. Column 1: source name; col. 2: type;
  col. 3: redshift; col. 4: luminosity distance; col. 5: scale.}
\begin{center}
\begin{tabular}{|l|c|c|c|c|}
\hline
Source&Type&z&D$_{\rm L}$&kpc/$^{''}$\\
 & & & &\\
\hline
&&&&\\
NGC\,1097&1&0.00424&18.0&0.086\\
MCG-5-23-16&2&0.008486&36.1&0.172\\
MRK\,1239&1.5&0.02883&124.5&0.570\\
NGC\,3783&1&0.009730&41.4&0.197\\
NGC\,5506&2&0.006181&26.2&0.126\\
NGC\,7469&1&0.016317&69.8&0.328\\
NGC\,7582&2&0.005254&22.3&0.107\\
&&&&\\
\hline
\end{tabular}
\end{center} 
\label{sample}
\end{table}

\section{New radio data}

Archival VLA/VLBA 
radio data of the aforementioned
sample were analysed in order to implement the
information available in the literature. To achieve an adequate
spatial resolution to resolve the nuclear
structure, we considered observations carried out when the VLA was in
one of the extended configuration (array A or B) at frequencies
ranging between 1.4 and 14.9 GHz, when available. We reduced datasets
either at frequencies lacking information or for which data presented in the
literature were not found satisfactory.\\
The data reduction was carried out following the standard procedures
for the VLA implemented in the NRAO AIPS package. Images were produced
after a few phase-only calibration iterations. 
In order to obtain accurate flux density at 1.4 GHz, it was necessary
to image several confusing sources falling within the primary beam.
Uncertainties in the
determination of flux density are
dominated by amplitude calibration errors, which
are between 3\% and 5\%, being worse at 14.9 GHz. 
The rms noise level (1$\sigma$)
on the image plane is usually below 0.1 mJy, being irrelevant for our
targets, with the exceptions of NGC\,1097 and MCG-5-23-16
where it is comparable to the amplitude calibration errors.\\
In the case of NGC\,1097, MRK\,1239 and NGC\,3783, for which no
information on their pc-scale structure could be found in the
literature, we reduced archival VLBA data. No archival VLBA data were found
for the sources MCG-5-23-16 and NGC\,7582. 
The Seyfert nuclei considered in this paper are
not bright enough to give detectable fringes. VLBI fringe
fitting was done on a bright source lying close to the
target, and off-beam phase referencing was performed to calibrate the phases of
the target source. 
In the case of NGC\,1097, no signal of the source 
was detected.
For MRK\,1239 and NGC\,3783, the structure was visible from the
first image with a 4$\sigma$ noise level. 
Then phase self-calibration with a solution interval of
30s was performed using the CLEAN component model. After a few
iterations we use natural weights to pick up possible extended
emission. No amplitude calibration was performed due to the weakness
of the targets.\\

\section{Results}

\subsection{Radio images}

\begin{table*}  
\caption{Radio observations and source parameters for the Seyfert
  galaxies. Column 1: source name; col. 2: observing frequency;
  col. 3: array configuration; col. 4: beam size; col. 5: observing
  date; col. 6: flux density; col. 7: luminosity; col. 8: largest
  angular size; col. 9: largest linear size; col. 10:
  morphology of the nuclear component (Un=unresolved, MR= marginally
  resolved; Double=two well-separated components); 
col. 11: morphology of the extended structure, when
  present. }
\begin{center}
\begin{tabular}{|l|c|c|c|c|c|c|c|c|c|c|}
\hline
Source&Freq.&conf&beam&Obs. date&Flux density&Log P&LAS&LLS&Morph.&Ext\\
      &GHz  &    &arcsec& &mJy&W/Hz&arcsec&pc& &\\
(1)&(2)&(3)&(4)&(5)&(6)&(7)&(8)&(9)&(10)&(11)\\
\hline
&&&&&&&&&\\
NGC\,1097&8.4&A&0.66$\times$0.25&Oct 31 1992&30.0&21.04&18&1550&Un&Ring\\
NGC\,1097&14.9&B&1.15$\times$0.45&Feb 24 2001&5.0&20.30&$<$0.5&$<48$ &Un&No\\
MCG-5-23-16&8.4&A&0.43$\times$0.23&Aug 21 1999&2.6&20.60&1.2&206&MR&No\\
MRK\,1239&1.4&A&1.54$\times$1.25&Jan 16 1993&60.0&23.04&$<$0.3&$<$171&Un&No\\
MRK\,1239&4.8&A&0.46$\times$0.40&Jul 31 1987&27.0&22.60&$<$0.1&$<$57&Un&No\\
MRK\,1239&8.4&A&0.33$\times$0.26&Jan 16 1993&15.0&22.44&$<$0.1&$<$57&Un&No\\
MRK\,1239&8.4&A&0.26$\times$0.22&Jul 15 1995&14.5&22.44&$<$0.1&$<$57&Un&No\\
MRK\,1239&14.9&B&0.97$\times$0.42&Dec 15 2003&9.0&22.22&$<$0.2&114&Un&No\\
MRK\,1239&1.6&VLBA&0.011$\times$0.004&May 09 2005&10.2&22.26&0.070&40&Double&No\\
NGC\,3783&1.6&VLBA&0.028$\times$0.008&Nov 30
1999&4.6&20.95&0.022&4.3&Un&No\\
NGC\,5506&1.4&A&1.71$\times$1.24&May 06 2002&324.0&22.42&4.0&500&Un&Halo\\
NGC\,5506&4.8&A&0.48$\times$0.37&May 06 2002&181.6&22.11&4.2&530&Un&Halo\\
NGC\,5506&8.4&A&0.27$\times$0.20&Nov 14 2000&111.6&21.96&3.8&480&Un&Halo\\
NGC\,5506&14.9&B&0.72$\times$0.43&Dec 05 1991&65.0&21.72&3.3&415&Un&Halo\\
NGC\,7469&8.4&A&0.22$\times$0.19&Feb 05 2006&31.4&22.26&3.1&1200&MR&Ring\\
NGC\,7469&14.9&A&0.14$\times$0.10&Sep 08 1999&10.6&21.71&0.1&33&MR&No\\
NGC\,7582&4.8&A&1.25$\times$0.32&Dec 12 2000&75.0&21.64&10&1070&Un&Ring\\
NGC\,7582&8.4&A&0.75$\times$0.19&Dec 12 2000&42.0&21.34&6&640&Un&Ring\\
&&&&&&&&&&\\
\hline
\end{tabular}
\end{center}
\label{observation}
\end{table*}

\begin{figure}
\begin{center}
\includegraphics{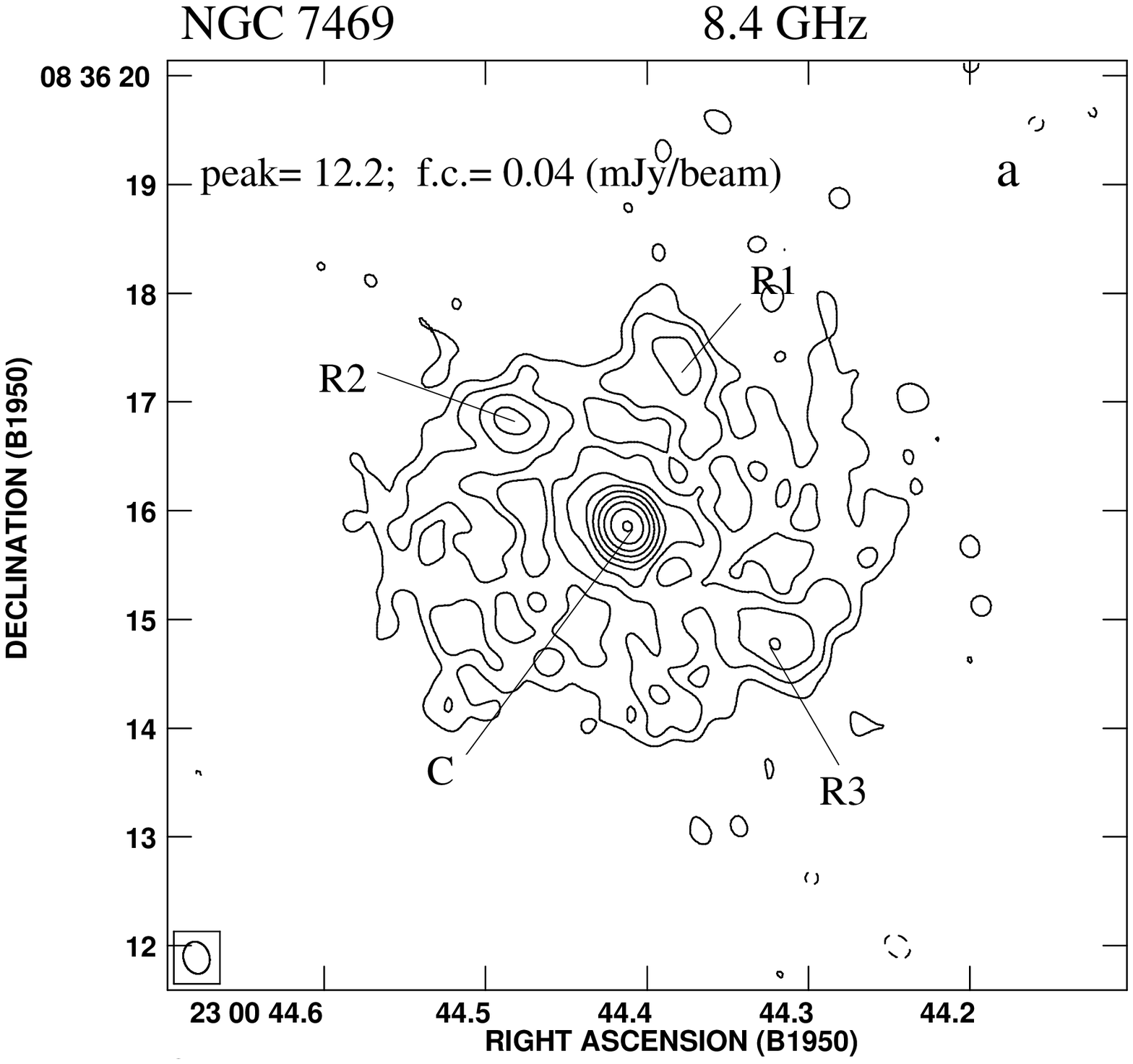}
\includegraphics{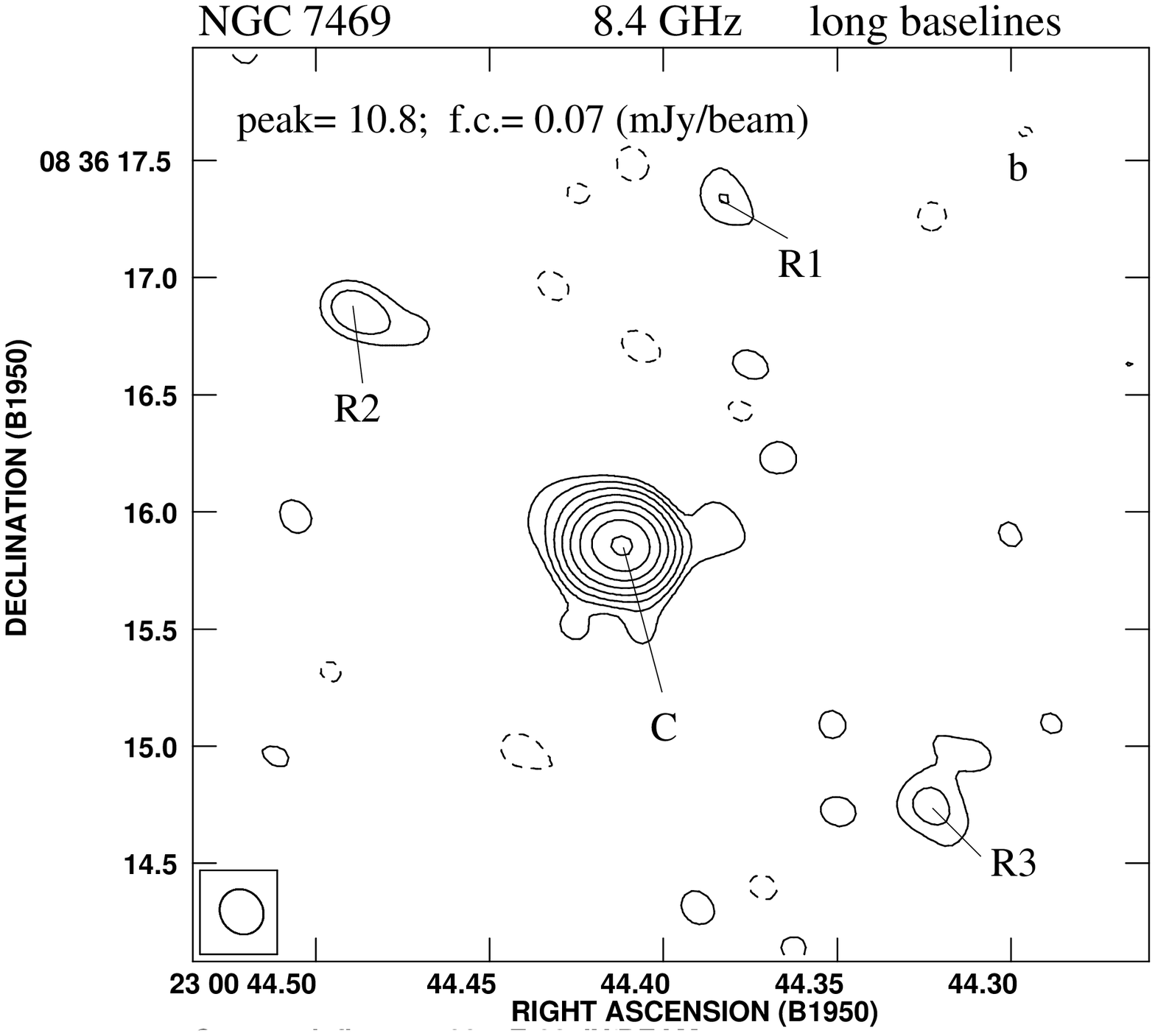}
\includegraphics{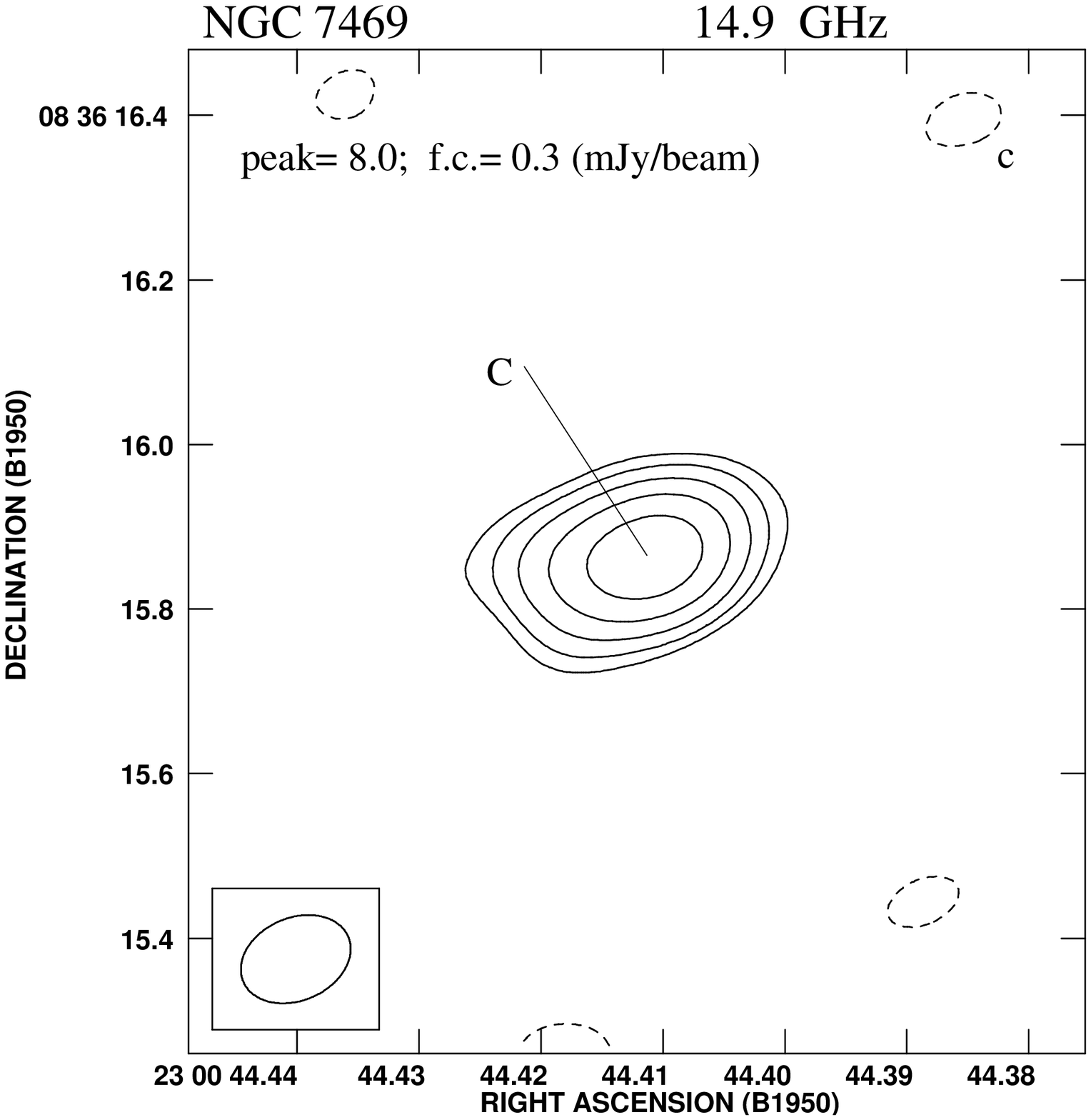}
\vspace{21.5cm}
\caption{VLA images of the central region of NGC\,7469 at 8.4 GHz ({\it top}),
    without the shortest baselines ({\it centre}), and at
    14.9 GHz ({\it bottom}). On each image we
    provide the observing frequency; the restoring beam, plotted on
    the bottom left corner; the peak flux density in
    mJy/beam; the first contour intensity ({\it f.c.} in mJy/beam),
    that is 3 times the off-source noise level; contour levels
    increase of a factor 2.}
\label{n7469}
\end{center}
\end{figure}

Full resolution VLA images of the sources discussed in this
paper are presented in 
Figs. \ref{n1097} 
to \ref{n7582}. 
In the case of NGC\,7469 and
NGC\,7582,  where an extended diffuse
emission is present around the central nucleus, an image without
the shortest baselines was produced in order to better describe the
compact components without the contamination from the low-surface
brightness extended emission (Figs. \ref{n7469}b and \ref{n7582}b).
For MRK\,1239 and NGC\,3783, VLBA images at 1.6 GHz are 
presented in Figs. \ref{m1239}c and \ref{n3783} respectively.\\
Flux density and
angular size were measured by means of the task JMFIT, which performs
a Gaussian fit on the image plane. In case of extended structures,
the flux
density was derived by TVSTAT, while the source size was measured from
the lowest contour on the image plane. 
The nucleus of these Seyfert galaxies is
unresolved or marginally resolved in VLA observations. We consider
marginally resolved those nuclei whose largest angular size (LAS)
is between 0.5 and 1 beam size at the best resolution, and we term
unresolved those whose LAS is smaller than half of the beam.\\
Source parameters, together with information on the observations, are
reported in Table \ref{observation}. In Table 3, we present
the observational parameters of the source components.\\

\subsection{Notes on individual sources}

\begin{table}
\begin{center}
\caption{Radio properties of the source
  components. Col. 1: source name; col. 2: source component, the name
refers to the label as reported in Figs. from \ref{n1097} to
\ref{n3783}; 
col. 3:
  observing frequency; col. 4: flux density; cols. 5 and 6:
  deconvolved major and minor axis.
$a$: values derived on VLBA images. }
\begin{tabular}{|l|c|c|c|c|c|}
\hline
Source&Comp.&Freq.&Flux&$\theta_{\rm max}$&$\theta_{\rm min}$\\
 & &GHz&mJy&arcsec&arcsec\\
(1)&(2)&(3)&(4)&(5)&(6)\\
\hline
&&&&&\\
NGC\,1097&C&8.4&4.0&$<$0.5& - \\
       &C&14.9&5.6&$<$0.5& -\\
       &R1&8.4&1.8&0.9&0.7\\
       &R2&8.4&0.8&1.7&1.3\\
       &R3&8.4&0.6&1.6&0.8\\
MCG-5-23-16&A&8.4&2.1&$<$0.1& - \\
           &B&8.4&0.5&$<$0.1& - \\
MRK\,1239&C&1.4&60.0&$<0.3$& - \\
       &C&5.0&26.8&$<$0.1& - \\
       &C&8.4&15.5&$<$0.1& - \\
       &C&14.9&9.3&$<$0.2& - \\
       &A$^{a}$&1.6&6.0&0.013&0.006\\
       &B$^{a}$&1.6&4.2&0.013&0.004\\
NGC\,3783&C$^{a}$&1.6&4.6&0.022&0.007\\
NCG\,5506&C&1.4&304.0&$<$1.0& - \\
       &C&4.8&136.0&$<$0.1& - \\
       &C&8.4&84.0&$<$0.08& - \\
       &C&14.9&58.0&$<$0.2& - \\
       &Wing&4.8&27&2.4&2.0\\
       &Wing&8.4&16&2.2&1.7\\
       &Halo&4.8&18&4.2&3.0\\
       &Halo&8.4&11&3.8&3.2\\  
NGC\,7469&C&8.4&14.8&$<$0.14& - \\
         &C&14.9&10.6&0.1&$<$0.05\\
         &R1&8.4&0.3&0.8&0.4\\
         &R2&8.4&0.6&0.9&0.7\\
         &R3&8.4&0.7&1.1&0.8\\
NGC\,7582&C&4.9&9.5&0.8&$<$0.1\\
         &C&8.4&6.9&0.4&0.2\\
         &R1&4.9&3.6&0.4&0.3\\
         &R1&8.4&2.1&0.3&0.3\\
         &R2&4.9&0.8&0.3&$<$0.1\\
         &R3&4.9&1.1&0.5&0.3\\
&&&&&\\
\hline
\end{tabular}
\end{center}
\label{components}
\end{table}

In this section we describe the characteristics of the Seyfert nuclei 
arising from the new images presented in this paper.\\

\subsubsection{NGC\,1097}

The new images at 8.4 and 14.9 GHz are presented in Fig. \ref{n1097},
and have a resolution of 0$^{\prime\prime}$.66$\times$0$^{\prime\prime}$.25 
and 1$^{\prime\prime}$.15$\times$0$^{\prime\prime}$.45 respectively.  
Fig. \ref{n1097}a shows for the first time the ring structure at 8.4 GHz.
Four star-forming regions, labelled R1, R2, R3 and R4
in Fig. \ref{n1097}a, are also present at 1.4 and 5 GHz
\citep[see e.g.][]{hummel87}. 
The central component appears
unresolved in both 8.4 and 14.9 GHz images,
giving an upper limit to its linear
size of $<0^{\prime\prime}.5$ ($<$43 pc). 
The spectral index is found inverted between
1.4 and 14.9 GHz. \\

\subsubsection{MCG-5-23-16}

The VLA image at 8.4 GHz (Fig. \ref{m-5}), with a resolution of
0$^{\prime\prime}$.43$\times$0$^{\prime\prime}$.23, 
is presented for the first time. 
The source is
 resolved in the north direction, suggesting the presence of a
 jet-like structure. 
The total 
radio spectrum shows a slight steepening at 8.4 GHz with a
spectral index $\alpha = 0.8 \pm 0.1$.\\

\subsubsection{MRK\,1239}

New 14.9-GHz VLA and 1.6-GHz VLBA images are presented in Figs. \ref{m1239}b,c
with resolution of
0$^{\prime\prime}$.97$\times$0$^{\prime\prime}$.42 and
0$^{\prime\prime}$.011$\times$0$^{\prime\prime}$.004 respectively. 
The central structure is unresolved in
our VLA images at all frequencies, giving an upper limit to the linear
size of $<$ 57 pc (Figs. \ref{m1239}a,b). 
The spectral index we obtain is $\alpha \sim 0.9 \pm 0.1$ 
without any evidence of the steepening reported by 
\citet{ulvestad95} (see Section 2.1.3), and in a better agreement with
what found by \citet{rush96}. 
It is worth noting that data reported by
\citet{rush96} refer to low resolution observations 
carried out when the VLA was in
D-configuration. The good agreement between these values and our
measurements suggests that no
extended emission on kpc scales is present. 
When observed with pc-scale resolution, the
nucleus of MRK\,1239 at 1.6 GHz 
is resolved in two components separated by
$\sim$ 50 mas ( $\sim$ 30 pc), with position angle of
40$^{\circ}$ (Fig. \ref{m1239}c). 
The lack of
multi-frequency observations with similar resolution does not allow us
to study the spectral index of these components.\\  
The flux density measured on the 
VLBA image at 1.6 GHz 
is 10.2 mJy, i.e. only 20\% of the VLA flux density at 1.6
GHz, obtained re-scaling the 1.4 GHz VLA flux density with the spectral
index computed between 1.4 and 4.8 GHz ($\alpha$ = 0.6). This indicates
that almost 80\% of the flux density measured with the VLA is
missing in the VLBA image. \\

\subsubsection{NGC\,3783}

The VLBA image at 1.6 GHz, with a resolution of
0$^{\prime\prime}$.028$\times$0$^{\prime\prime}$.008
(Fig. \ref{n3783}), 
is presented for the first time.  
In this image, the nucleus is unresolved
with a linear size $<$ 4
pc. The total flux density measured on the VLBA image is 4.6 mJy,
that represents only a 20\% of the flux density of the unresolved
component in VLA observations \citep{unger86}. This indicates that the
majority ($\sim$80\%) of the flux measured with the VLA is missing in
the VLBA image.\\

\subsubsection{NGC\,5506}

New VLA images at 4.8 and 8.4 GHz with resolution of
0$^{\prime\prime}$.48$\times$0$^{\prime\prime}$.37 
and 0$^{\prime\prime}$.27$\times$0$^{\prime\prime}$.20 respectively, 
are shown in Fig. \ref{n5506}a,b.  
The source 
displays an unresolved
central component with a diffuse wing-like emission extending mainly to the
north-west and east part of the nucleus. 
The high dynamic range, namely the ratio between the
peak flux density and 1$\sigma$ noise level, of our images
allows us to detect also at 8.4 GHz
the extended low-surface
brightness halo (Figs. \ref{n5506}a,b). This has a diameter of
$\sim$ 2$^{\prime\prime}$.75 ($\sim$ 350 pc) enshrouding the central
features. The nucleus is unresolved in our VLA images, giving an upper
limit of $<$0$^{\prime\prime}$.08 ($<$10 pc).
The nucleus accounts for the
majority (75\%) of the radio emission. Its spectral index is $\alpha \sim 0.8
\pm 0.1$. 
The low-surface brightness halo and the extended wing-like structures
have 
a slightly steeper spectral index
$\alpha = 0.9 \pm 0.1$.\\
A comparison between simultaneous MERLIN and EVN observations at 18 and 6 cm
clearly indicates that almost 43\% of the flux density detected by
MERLIN cannot be recovered in EVN images \citep{middelberg04}. In
fact, a diffuse emission, detected by MERLIN observations, is not seen
by the EVN, as the EVN is insensitive to structures larger than 35 and 11
mas at 18 and 6 cm, respectively.\\

\begin{figure}
\begin{center}
\includegraphics{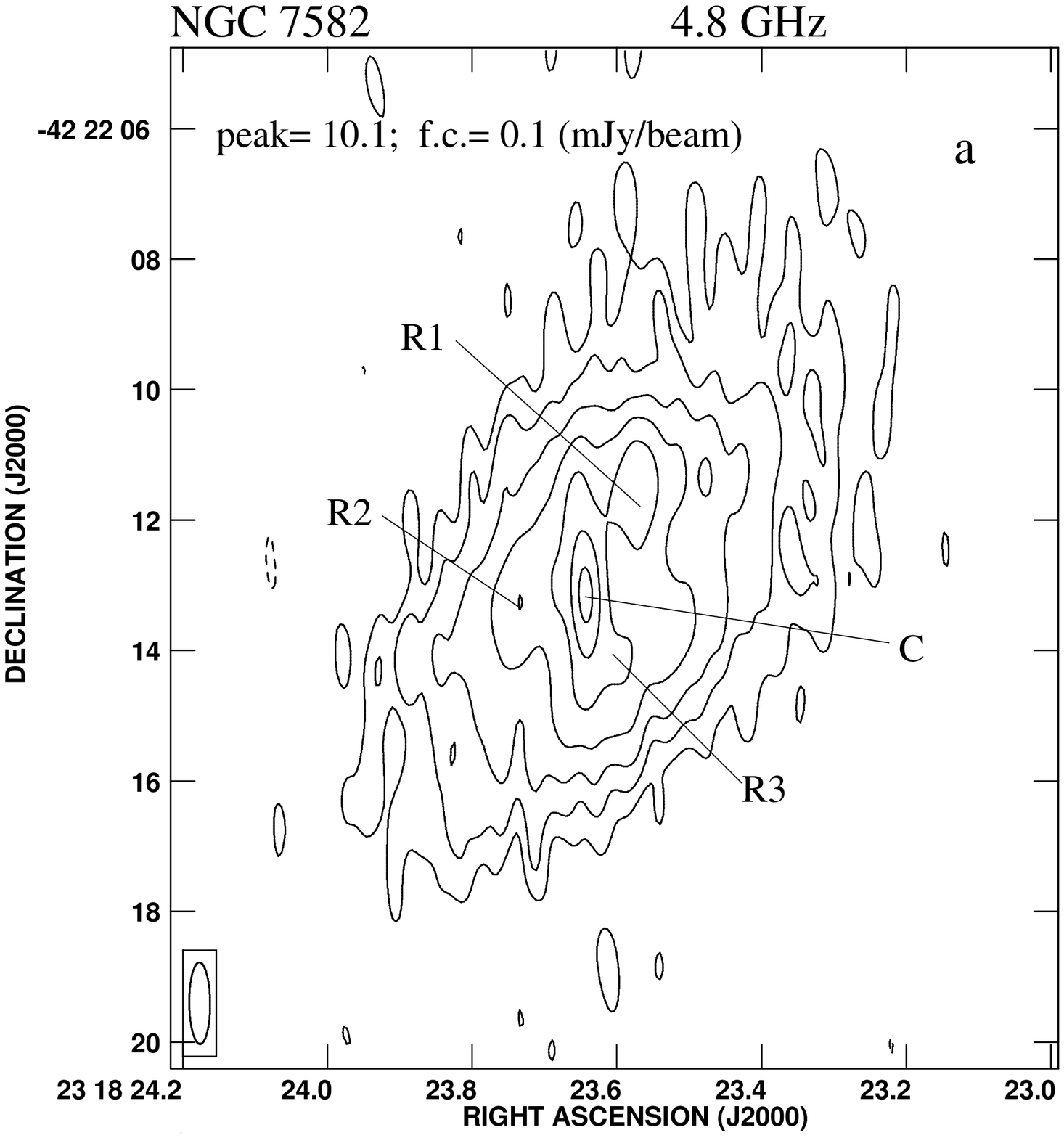}
\includegraphics{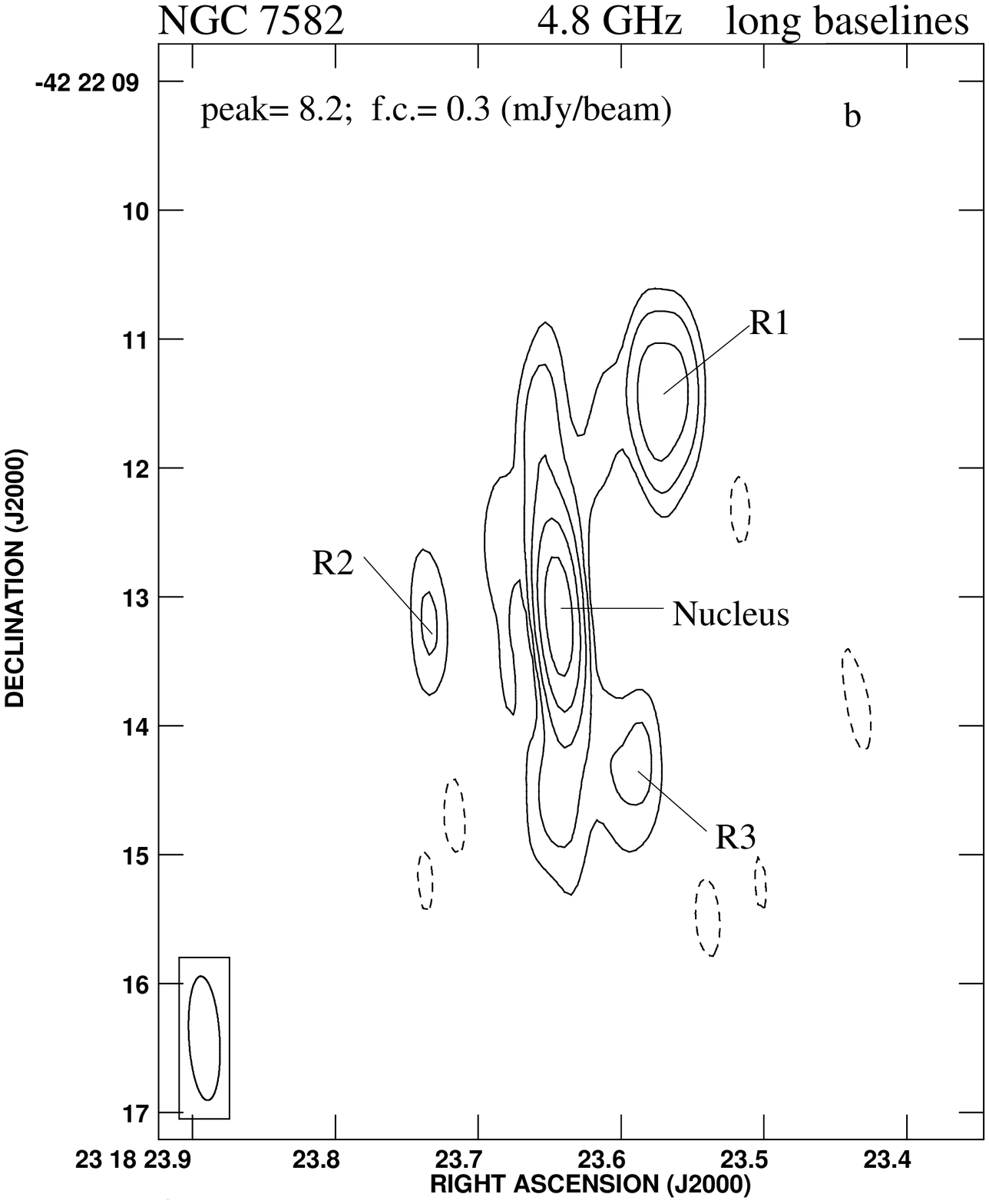}
\includegraphics{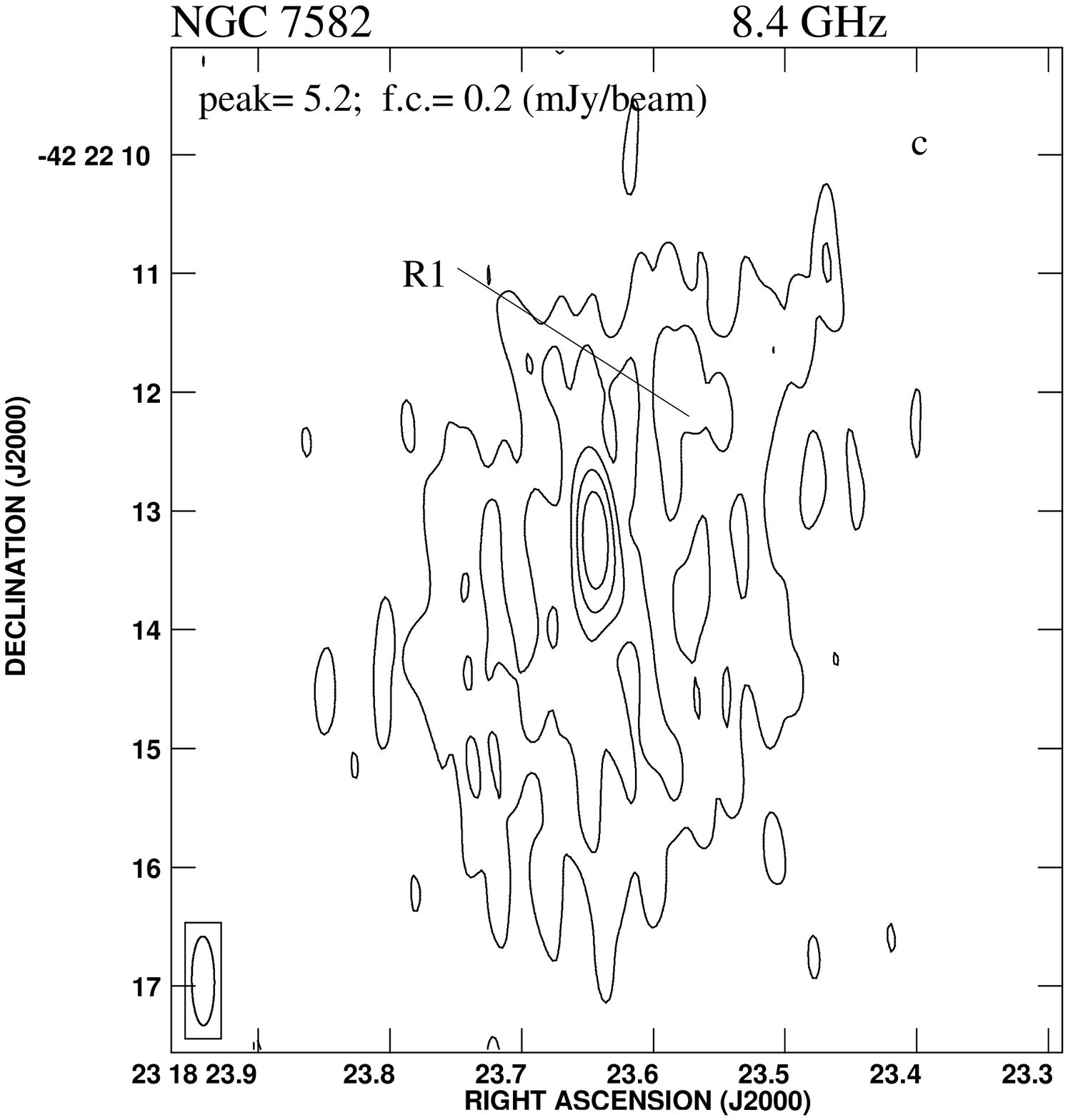}
\vspace{21.5cm}
\caption{VLA images of the central region of NGC\,7582 at 
4.8 GHz ({\it top}), without the shortest baselines ({\it centre}), and 
at 8.4 GHz ({\it bottom}). 
On each image we
    provide the observing frequency; the restoring beam, plotted on
    the bottom left corner; the peak flux density in
    mJy/beam; the first contour intensity ({\it f.c.} in mJy/beam),
    that is 3 times the off-source noise level; contour levels
    increase of a factor 2.}
\label{n7582}
\end{center}
\end{figure}

\subsubsection{NGC\,7469}

We produced new VLA image at 8.4 GHz with a resolution of
0$^{\prime\prime}$.2$\times$0$^{\prime\prime}$.18 (Fig. \ref{n7469}b) 
obtained without the
shortest ($<35$ k$\lambda$) baselines in order to pinpoint the compact
central structure and to reduce the  
contamination from the surrounding ring.  
The radio emission is dominated by an unresolved central component
surrounded by
a ring of star-forming regions with a
diameter of about 3$^{''}$.7 ($\sim$ 1.2 kpc). 
In our new image at 14.9 GHz (Fig. \ref{n7469}c), the nucleus is
elongated in the east-west direction.
The spectral index is $\alpha \sim 0.5 \pm 0.1$. 
The elongation is
in agreement with that shown by MERLIN 
\citep{alberdi06} and VLBI \citep{lonsdale03} images 
where the nucleus clearly shows a
core-jet structure elongated in the same direction. 
In the diffuse extranuclear emission,
three clumps of star-forming regions, 
labelled R1, R2, and R3 in Figs. \ref{n7469}a,b, are clearly
visible at 8.4 GHz.\\
A comparison between MERLIN and VLBA
images clearly indicates that almost 50\% of the flux density
detected by MERLIN \citep[FWHM $\sim$
0$^{\prime\prime}$.15,][]{thean01} 
is missing in the VLBA image at the same frequency
\citep{lonsdale03}. \\

\begin{figure}
\begin{center}
\includegraphics{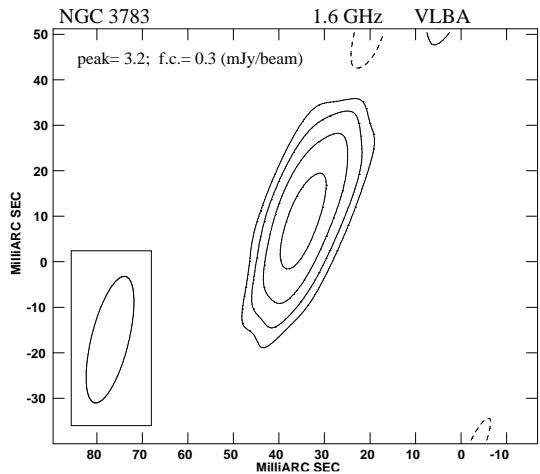}
\vspace{7.0cm}
\caption{VLBA image at 1.6 GHz of the central region of NGC\,3783. 
On the image we
    provide the observing frequency; the restoring beam, plotted on
    the bottom left corner; the peak flux density in
    mJy/beam; the first contour intensity ({\it f.c.} in mJy/beam),
    that is 3 times the off-source noise level; contour levels
    increase of a factor 2.}
\label{n3783}
\end{center}
\end{figure}

\subsubsection{NGC\,7582}

We produced new VLA image at 4.8 GHz with a resolution of
0$^{\prime\prime}$.94$\times$0$^{\prime\prime}$.24 (Fig. \ref{n7582}b)
obtained without the
shortest ($<45$ k$\lambda$) baselines in order to pinpoint the compact
central structure and to reduce the  
contamination from the surrounding extended emission. 
The radio emission is dominated by 
an unresolved central
component surrounded by diffuse emission related to star-forming
regions (Fig. \ref{n7582}), also detected in the optical. 
The nucleus, 
with a spectral index $\alpha \sim 0.6 \pm
0.1$, is unresolved at all frequencies, giving an
upper limit of about 0$^{\prime\prime}$.4 ($<$40 pc) to its linear
size. The extended emission has a maximum diameter of about
9$^{''}$ ($\sim$ 950 pc) and a spectral index $\alpha \sim 1.0
\pm 0.1$, that is steeper than the value ($\alpha \sim 0.7$) found
by \citet{morganti99}. The discrepancy may arise from the different
resolution of the observations taken into account. 
Three compact regions, 
labelled R1, R2, R3 in Figs. \ref{n7582}a and \ref{n7582}b, 
are present in the
diffuse emission at 4.8 GHz. 
For these clumps, near-infrared counterpart was found
by Fernandez-Ontiveros (2009, in preparation) 
suggesting that they are star-forming regions. 
Component R1 is visible also at 8.4 GHz, with a
flux density of $\sim$2.1 mJy, and a spectral index $\alpha \sim
0.9 \pm 0.2$, that is a little flatter than the spectral index of the
whole extended emission.\\   

\section{Discussion}

\subsection{The nuclear radio emission}

In general, the nucleus of these Seyfert galaxies 
is unresolved at the resolution
provided by the VLA in its larger configurations (A and B arrays), 
giving an upper limit to the linear size between
about 20 and 120 pc. 
Only MCG-5-23-16 and NGC\,7469
are marginally resolved at the highest resolution (FWHM $\sim$ 0.2
arcsec). In both cases the central component
is elongated suggesting a core-jet morphology.\\
For all the sources we find that the spectral index of the nucleus 
ranges between 0.5 and 0.9, with
the only exception of NGC\,1097 where the spectrum is inverted.\\
When observed with pc-scale resolution, most sources
display a resolved structure. This
is the case of MRK\,1239 (Fig. \ref{m1239}c), 
NGC\,5506 \citep{middelberg04},
and NGC\,7469 \citep{lonsdale03}. 
MRK\,1239 shows a well defined double morphology with a total linear size of 40
pc. The lack of multi-frequency information does not allow us to study
the spectral index distribution. 
NGC\,5506 and NGC\,7469 are characterised by
several compact components resembling a core-jet structure.
In NGC\,5506, the
availability of multi-frequency observations singles out one of the
components with a flat spectrum \citep{middelberg04}.\\
NGC\,3783 is still unresolved at
pc-scale resolution (Fig. \ref{n3783}),
and an upper limit to the linear
size $<$ 4 pc is estimated. \\
For the sources with VLBA information, 
MRK\,1239, NGC\,3783, NGC\,5506, and NGC\,7469,
an estimate of the brightness temperature $T_{\rm B}$, the
equipartition magnetic field $H_{\rm eq}$ and minimum energy density
$u_{\rm min}$ is provided. We do not compute physical parameters for
the sources unresolved in VLA images because no tight constraints on
their values can be derived. To compute the physical parameters of 
MRK\,1239 and NGC\,3783 we
make use of the values at 1.6 GHz reported in Table 3. 
In the case of NGC\,5506 and
NGC\,7469 we compute the physical parameters making use of the values
reported in \citet{middelberg04} and \citet{lonsdale03},
respectively.
The brightness temperature is computed by:

\begin{displaymath}
T_{\rm B} = \frac{S(\nu)}{2k \, \theta_{\rm maj} \theta_{\rm min}}
\left( \frac{c}{\nu} \right)^{2}  
\label{t_bright}
\end{displaymath}

\noindent where $S$($\nu$) is the flux density at the frequency $\nu$,
$\theta_{\rm maj}$ and $\theta_{\rm min}$ are the source
major and minor axis, 
$k$ is the Boltzmann constant, and $c$ the speed of light. 
For the magnetic
field and energy density we follow
standard
formulae \citep{pacho70}, assuming equipartition condition between
radiating particles and magnetic field. The equipartition magnetic
field is

\begin{displaymath} 
H_{\rm eq} = \sqrt{ \frac{24}{7} \pi u_{\min}}
\label{h_equi}
\end{displaymath}

\noindent and the minimum energy density $u_{\rm min}$ is

\begin{displaymath}
u_{\rm min} = 1.04 \cdot 10^{-23} \cdot \left( \frac{L}{V}\right)^{4/7}
\label{umin}
\end{displaymath}

\noindent where $L$ is the synchrotron luminosity in Watt, 
and $V$ the volume of
the emitting region in pc$^{3}$.
We assume that the volume of the emitting region is a prolate
spheroid

\begin{displaymath}
V = \frac{\pi}{6} d_{\rm maj} d_{\rm min}^{2}  
\end{displaymath}

\noindent where $d_{\rm maj}$ and $d_{\rm min}$ are the source linear
sizes. We assume a filling factor of unity, i.e. the
source volume is fully and homogeneously filled by relativistic
plasma. Furthermore, proton and electron energy densities are assumed to be
equal.
For the Seyfert nuclei with VLBI information, we obtain $u_{\rm
min}$ of the order of 10$^{-6}$ -- 10$^{-7}$ erg/cm$^{3}$,
$H_{\rm eq} \sim $1.5 -- 10 mG 
and $T_{\rm B}$ of about 10$^{7}$ -- 10$^{8}$ K. These
values are comparable to those derived in the nucleus of other Seyfert
galaxies \citep[e.g.][]{ulvestad84, kukula99}.

\subsection{The off-nucleus radio emission}

Of the seven Seyfert nuclei studied, four 
(NGC\,1097, NGC\,5506, NGC\,7469 and NGC\,7582) are surrounded by
diffuse 
emission extending on kpc scale. 
In NGC\,5506, 
\citet{wehrle87} suggested that its extended 0.35-kpc halo may be due to
either a radio plasma bubble expanding from the AGN, or
a magnetically dominated coronal arc. 
Another possibility is that
it is produced by free-free emission from the
ionised gas heated by the AGN. However, the steep spectral
index $\alpha = 0.9$ makes the interpretation unlikely, as
thermal emission would produce a flatter spectral
index. Furthermore, 
the extended emission does not seem to be related to starburst
activity, as no evidence for this is seeing in high spatial
resolution IR images up to 20 $\mu$m \citep{prieto09,reunanen09}. \\
In NGC\,1097, NGC\,7469 and NGC\,7582,
the extended emission is resolved in knots (Table 3), 
each having a counterpart
star-forming region in high spatial resolution IR images
\citep{reunanen09}. The analysis of the SED of the individual
star-forming clumps in these galaxies, from radio to UV, is showing
that these regions are young super-massive stellar clusters, analogues
to those seen in the starburst nucleus of NGC\,253
\citep{ferna09}. In radio, the spectral
index is relatively steep, $\sim$ 0.4 - 0.9 in NGC\,1097
\citep{hummel87} and in NGC\,7469
\citep{alberdi06}, and $\sim$0.9 in region R1 of NGC\,7582 (no
sufficient information is available for the other knots of NGC\,7582
to derive the spectral index), suggesting a dominant non-thermal
origin, most probably from supernovae events in these clusters. Not all
the IR star-forming regions are detected in radio, suggesting
different evolution phase of the clusters within the same star-forming
ring (Fernandez-Ontiveros et al. 2009, in preparation). \\

\subsection{The missing flux density}

\subsubsection{Observational limitations}

An interesting characteristic shown by the Seyfert 
nuclei with VLBI information, namely 
MRK\,1239, NGC\,3783, NGC\,5506 and NGC\,7469,
is that their
pc-scale flux density is significantly lower than that measured on VLA
images even in the presence of unresolved structures. 
In MRK\,1239 and
NGC\,3783, the flux density derived from VLBA data accounts only for 20\%
of that measured on VLA images. 
This suggests that a significant fraction of the radio emission is not
concentrated in a compact component at 
the centre, but it spreads
over a region that is slightly larger than 0$^{\prime\prime}$.1, i.e. the largest angular scale
detectable by the VLBA at 1.6 GHz, and smaller than
the resolution of VLA data at the same
frequency, that is $<0^{\prime\prime}$.3 for MRK\,1239
(see Table \ref{observation}) and $<0^{\prime\prime}$.5 for
NGC\,3783 \citep{ulvestad84}. 
A strong support to this interpretation comes from observations with different
resolution of the nucleus of NGC\,7469. In MERLIN
observations at 1.6 GHz \citep{alberdi06} 
the nucleus is slightly resolved on the east-west direction, giving an
upper limit to its angular size of about 0$^{\prime\prime}$.15. The east-west
elongation is confirmed by pc-scale resolution VLBI observations
\citep{lonsdale03}, where the nucleus is resolved in 5 
components spread over an area of about 0$^{\prime\prime}$.17,
i.e. comparable to
that derived from MERLIN data.
However, the flux density recovered from
VLBI data accounts only for $\sim$50\% of that measured on MERLIN
images. A similar result was found in NGC\,5506 by
\citet{middelberg04}, where EVN observations of the nucleus
could recover only 43\%
of the flux density measured by MERLIN. In this case, an extended,
low-surface brightness
region, detected by MERLIN, is not seen by the EVN, as it is
insensitive to structures larger than 35 mas at 18 cm, and 11 mas at 6 cm.\\ 
From these results we can argue that in the case of NGC\,5506 and
NGC\,7469,
the missing flux
density on VLBI image is likely due to diffuse, steep-spectrum
low-surface brightness
emission that is undetectable by VLBI observations either due to
sensitivity limitation or because the structure is larger than the
maximum scale detectable by the interferometer, or a combination of
both. A similar result was found in the case of the Seyfert 
NGC\,1068 where \citet{gallimore04} noted that 
deep VLBA observations could detect a flux density almost 50\% higher
than that of lower sensitivity VLBA observations by
\citet{roy98}, but still the total flux density measured
on images with lower spatial resolution could not be recovered. \\

\subsubsection{Steep-spectrum versus flat-spectrum Seyfert nuclei}

The fact that in Seyfert nuclei the pc-scale radio emission does not
account for all the flux density measured at lower resolution, even in
the presence of an unresolved component, was already noted by
\citet{sadler95} by studying a sample of 22 nearby Seyferts with
the PTI. They 
suggested that in Seyfert galaxies the radio
emission is more diffuse and less centrally concentrated than in
elliptical radio galaxies. A similar behaviour was also found by
\citet{lal04} by comparing simultaneous 
VLA and VLBI
observations of a sample of Seyfert galaxies. They found that
in $\sim$ 60\% of the sources more than half of the VLA flux density
of the unresolved component is
missing in VLBI images. \\
However, evidence of undetected flux density on pc-scale is not
a characteristic common to all Seyfert nuclei. For example, it has
been found that in flat-spectrum Seyfert nuclei essentially all the
emission on arcsecond scale is present on VLBI scales
\citep{anderson04, anderson05}, indicating  that all the radio
emission is concentrated in the central component. 
On the other hand, in Seyfert nuclei characterised by a steep
spectrum, the flux density recovered on mas-scales is usually
significantly lower than that expected from arcsecond-scale images,
suggesting that the radio emission is not centrally
concentrated, but it is diffuse on a larger region. A noticeable
example of a steep-spectrum Seyfert nucleus 
where the flux density almost disappears
moving from arcsec to milli-arcsec scales is represented by
NGC\,4151, where only 8\% of the flux density from the central
component could be recovered by VLBI observations \citep{pedlar93,
ulvestad98}, while a large fraction of the jet emission is undetected.
This suggests that in Seyfert nuclei
unresolved on arcsec scale but with different spectral index
properties, the dominant radio emission may originate from different
features: from the central core in flat-spectrum objects, from
extended features, like jets, in steep-spectrum nuclei.\\

\subsubsection{Thermal or non-thermal origin?}

So far, the nature of the missing flux
density has not been investigated in detail. 
This diffuse emission may be 
due to either thermal
emission from ionised gas heated by the AGN, or by nuclear star
forming regions, or non-thermal
radiation from the AGN itself. \\
To investigate a possible free-free origin of the missing flux
we compute the electron density $n_{e}$ that the
ionised gas must have in order to emit the ``missing'' flux
density:\\

\begin{equation} 
n_{e}^{2} = 1.84 \times 10^{41} \left( \frac{T}{10^{4} {\rm K}}
\right)^{1/2}D_{\rm   L}^{2}S(\nu)V^{-1}g_{\rm ff}^{-1}   
\label{free}
\end{equation}

\noindent where $D_{\rm L}$ is the luminosity distance, $T$ is the gas
temperature in units of 10$^{4}$ K, and $g_{\rm ff}$ the Gaunt factor.  
In Eq. \ref{free} we assumed that electrons and protons
have the same density. At radio frequency, $g_{\rm ff} \sim 17.7+{\rm ln}(T^{3/2}/
\nu)$ \citep[e.g.][]{pacho70}. We assume the gas temperature $T
\sim 10^{4} -
10^{6}$ K.  
The missing flux is calculated as the difference between the VLA and
VLBI flux densities.
In the case of MRK\,1239 and NGC\,3783, the missing flux
at 1.6 GHz is $\sim$ 50 and 18 mJy, respectively. 
If we consider these parameters in Eq. \ref{free},
we obtain an upper limit to the electron density $n_{e}
\sim 10^{3} - 10^{4} {\rm cm}^{-3}$, over a spherical volume of 0.1 arcsec 
(i.e. 57 and 20 pc
for MRK\,1239 and NGC\,3783 respectively) in
diameter.
This value is the largest angular size detectable from the
VLBA, (i.e. the VLBA at 1.6 GHz is insensitive to structures larger
than 0$^{\prime\prime}$.1) and it represents, therefore, 
a lower limit to the source size. 
Such a dense gas would completely absorb the
synchrotron emission arising from the embedded AGN:

\begin{equation}
S_{\rm obs}(\nu) = S_{0}(\nu)e^{- \mu(\nu) l}
\label{absorption}
\end{equation}

\noindent  
where $S_{\rm obs}$ and $S_{0} (\nu)$ are the observed and 
the intrinsic flux density at the
frequency $\nu$ respectively, $l$ 
is the width of the ionised region, and $\mu(\nu)$ the
absorption coefficient at the frequency $\nu$ \citep{rybicki79}. 
At radio wavelengths

\begin{displaymath} 
\mu (\nu) \sim 0.018 T^{-3/2} n_{e}^{2} \nu^{-2} g_{\rm ff}.
\end{displaymath}

\begin{figure}
\begin{center}
\includegraphics{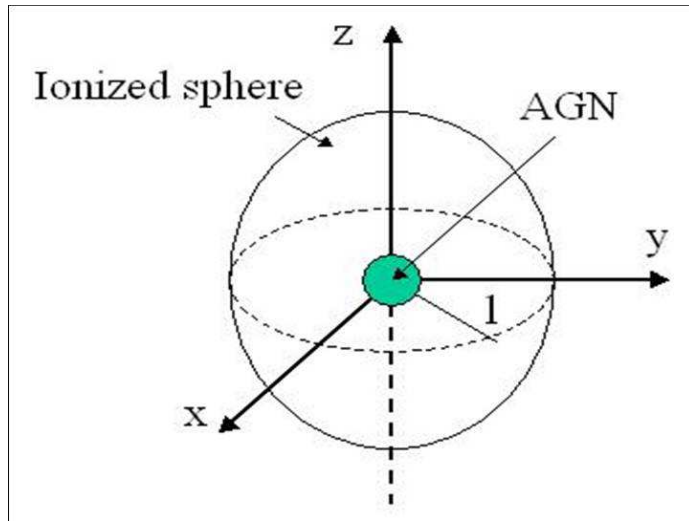}
\vspace{7.5cm}
\caption{The assumed geometry for the ionised gas heated by the
AGN. The AGN is located in the centre of the ionised sphere with an angular
diameter of 0.1 arcsec, that corresponds to the largest region
detectable by the VLBA at 1.6 GHz.}
\label{sfera}
\end{center}
\end{figure}

\noindent We consider that the pc-scale structure found in the VLBA
images (Figs. \ref{m1239}c and \ref{n3783}) arises from the central AGN
and the observed flux density is 10 mJy and 5 mJy for MRK\,1239 and
NGC\,3783 respectively. We assume that the AGN is  
located in the centre of the ionised region, i.e. $l$ is
equal to the radius of the ionised sphere (Fig. \ref{sfera}) that accounts for
28.5 pc and 10 pc in the case of MRK\,1239 and NGC\,3783,
respectively. With this parameters in Eq. \ref{absorption} 
we obtain an intrinsic flux density arising from the central AGN that
is much higher than a few thousands of Jy.
This suggests that the ``missing'' flux is not free-free from gas
ionised by the central AGN, and a non-thermal AGN-related origin is
more plausible.\\
Another possibility may be that the
ionised gas is related to an 
H$_{\rm II}$ region located in projection behind the AGN, or that
the gas around the AGN is not ionised uniformly.
In this way, our line of sight to the central AGN does not pass
through the ionised gas, thus avoiding any absorption of the AGN
radiation.\\
However, a strong support to the non-thermal 
synchrotron origin of the diffuse
emission comes from the comparison, when possible, between the
spectral index distribution in low and spatial resolution images. For
example, in the case of NGC\,5506, the unresolved nucleus in VLA data
has a steep spectrum with a mean value $\alpha \sim 0.8$, while at
pc-scale resolution \citep{sadler95} the spectrum has a convex shape
with the peak occurring around 3.5 GHz. This suggests that a
significant fraction of steep-spectrum, low-surface brightness
emission is present in the nuclear region, but it cannot be detected
in VLBI observations likely due to observational limitations. \\
As in the case of NGC\,4151 \citep{ulvestad98}, the steep emission
may arise from a jet, that may be distorted and/or disrupted by the
dense ambient medium. 
For example on arcsecond scale, NGC\,4151 
displays a compact core and
two collimated jets \citep{pedlar93,kukula95}. When observed with higher
resolution,  
the compact core is resolved in several knots forming a
jet which is not aligned with the arcsec-scale jet \citep{ulvestad98}.
The misalignment often found between the radio
jets on milli-arcsecond scale and those on larger scale
\citep{ulvestad98} may be an
evidence of the interaction between the radio jet and the
environment. In this scenario it is also possible that, for some reason, 
flat-spectrum Seyfert nuclei are not able to develop a radio jet, and
this causing that
all their emission is produced in the central compact component.\\

\section{Conclusions}

We presented the analysis of multi-frequency archival VLA
and VLBA
data of 7 close and bright Seyfert nuclei. 
The conclusions we can draw from this investigation are:\\

\begin{itemize}

\item At VLA resolution, FWHM $\sim 0^{\prime\prime}.1$, the nucleus
of the Seyfert galaxies is unresolved, with the only exception of
MGC-5-23-16 and NGC\,7469 which show 
a core-jet structure. At VLBA
resolution, equivalent to a few parsecs, 
the nucleus of MRK\,1239 is resolved in two components,
while in NGC\,3783 it is still unresolved. \\

\item The Seyfert galaxies in this study with known circumnuclear
star-forming regions in the IR, namely NGC\,1097, NGC\,7469 and
NGC\,7582 \citep{reunanen09} present a radio counterpart for a few
($\sim$10\%)
of these regions. Most are not detected in radio, indicating that
those detected may be characterised by higher supernovae rate. Their
steep radio spectral index is in line with this idea. 
Conversely, none of the other Seyfert nuclei present any circumnuclear
radio emission.
The only
exception is NGC\,5506 that shows a radio halo surrounding the nucleus. \\

\item A comparison between arcsecond and milli-arsecond resolution
in MRK\,1239 and NGC\,3783, 
pointed out that almost 80\% of the radio flux density detected in VLA
observations is not recovered at pc-scale resolution. This suggests
the presence of a diffuse component on scales of a few tens of
parsecs, undetected with the VLBA. Similar situation is found in
NGC\,1068, NGC\,4151, NGC\,5506 and NGC\,7469. The
nature of this ``missing'' flux components is likely due to
synchrotron AGN-related emission. This difference between the flux
density measured on arcsecond and milli-arcsecond resolution images
is not found in the case of elliptical radio galaxies, but
it appears to be a common phenomenon in Seyfert galaxies with
a steep spectrum, mostly hosted in spirals. 
If of synchrotron origin, this emission may be
spilt off from a jet that may get easily distorted and/or disrupted
by the dense interstellar medium in the nucleus of spirals.\\ 

\item A comparison between Seyfert nuclei with different spectral
properties points out that in flat-spectrum nuclei, almost all the
flux density is recovered on milli-arcsecond scale. This indicates that
in flat-spectrum objects the radio emission is essentially
concentrated in the compact core, without evidence of jet-like
structure even on milli-arcsecond scale, while
in steep-spectrum objects a significant fraction of the radio emission
arises from low-surface brightness, extended features.  
 
\end{itemize}

\section*{Acknowledgements}

We thank the anonymous referee for carefully reading the manuscript 
and valuable suggestions. 
The National Radio Astronomy Observatory is a facility of the National
Science Foundation operated under cooperative agreement by Associated
Universities, Inc. This work has made use of the NASA/IPAC
Extragalactic Database NED which is operated by the JPL, Californian
Institute of Technology, under contract with the National Aeronautics
and Space Administration.

\end{document}